\documentclass[%
 reprint,
superscriptaddress,
 amsmath,amssymb,
 aps,
]{revtex4-2}

\usepackage{graphicx}
\usepackage{dcolumn}
\usepackage{bm}
\usepackage{color}

\usepackage{algorithm}
\usepackage{algpseudocode}
\usepackage[hidelinks]{hyperref}
\usepackage{physics}
\usepackage{mathtools}
\usepackage{csquotes}

\usepackage{xcolor} 
\usepackage[columnwise]{lineno}

\newcommand{\red}[1]{{#1}}

\begin{document}
\preprint{APS/123-QED}

\title{Efficient Convex Optimization for Bosonic State Tomography}

\author{Shengyong Li}
\thanks{These authors contributed equally to this work.}
 \affiliation{Department of Automation, Tsinghua University, Beijing 100084, China.}
 
\author{Yanjin Yue}%
\thanks{These authors contributed equally to this work.}
\affiliation{%
 School of Physics, Sun Yat-sen University, Guangzhou 510275, China.
}%

\author{Ying Hu}
\affiliation{
Department of Physics and Synergetic Innovation Center for Quantum Effects and Applications, Hunan Normal University, Changsha 410081, China.
}

\author{Rui-Yang Gong}%
\affiliation{%
 School of Physics, Sun Yat-sen University, Guangzhou 510275, China.
}%

\author{Qianchuan Zhao}
 \affiliation{Department of Automation, Tsinghua University, Beijing 100084, China.}

\author{Zhihui Peng}
\affiliation{
Department of Physics and Synergetic Innovation Center for Quantum Effects and Applications, Hunan Normal University, Changsha 410081, China.
}

\author{Hou Ian}
\affiliation{Institute of Applied Physics and Materials Engineering, University of Macau, Macau S.A.R., China.}

\author{Pengtao Song}%
\email{ptsong@xjtu.edu.cn}
\affiliation{School of Automation Science and Engineering, Xi’an Jiaotong University, Xi’an, 710049, China.}

\author{Ze-Liang Xiang}%
\email{xiangzliang@mail.sysu.edu.cn}
\affiliation{%
 School of Physics, Sun Yat-sen University, Guangzhou 510275, China.
}%

 \author{Jing Zhang}%
\email{zhangjing2022@xjtu.edu.cn}
\affiliation{School of Automation Science and Engineering, Xi’an Jiaotong University, Xi’an, 710049, China.}
\affiliation{MOE Key Lab for Intelligent Networks and Network Security, Xi’an Jiaotong University, Xi’an 710049, China.}




\date{\today}

\begin{abstract}
Quantum states encoded in electromagnetic fields, also known as bosonic states, have been widely applied in quantum sensing, quantum communication, and quantum error correction. 
Accurate characterization is therefore essential yet difficult when states cannot be reconstructed with sparse Pauli measurements. Tomography must work with dense measurement bases, high-dimensional Hilbert spaces, and often sample-based data. However, existing convex optimization-based techniques are not efficient enough and scale poorly when extended to large and multi-mode systems.
In this work, we explore convex optimization as an effective framework to address problems in bosonic state tomography, introducing three techniques to enhance efficiency and scalability: efficient displacement operator computation, Hilbert space truncation, and stochastic convex optimization, which mitigate common limitations of existing approaches. Then we propose a sample-based, convex maximum-likelihood estimation (MLE) method specifically designed for flying mode tomography. Numerical simulations of flying four-mode and nine-mode problems demonstrate the accuracy and practicality of our methods. 
This method provides practical tools for reliable bosonic mode quantum state reconstruction in high-dimensional and multi-mode systems.

\end{abstract}

\maketitle


\section{\label{sec:intro}INTRODUCTION}

Quantum state tomography~\cite{PhysRevLett.105.150401, PhysRevLett.105.250403} (QST) is widely used in recent experiments to reconstruct quantum states from measurement data. It serves as the foundation for various quantum system identification tasks~\cite{PhysRevLett.108.080502}, such as Hamiltonian identification~\cite{PhysRevLett.113.080401, 8022944, 9026783, HOU2017863} and quantum process tomography~\cite{PhysRevLett.90.193601, PhysRevA.77.032322, doi:10.1126/science.1162086}. As such, quantum state tomography plays a crucial role in quantum sensing, quantum computation, and quantum communication~\cite{Lange2023adaptivequantum,Tripathi2025, Yang2025}.
Unlike many classical system identification problems, QST involves reconstructing the quantum state from samples that are drawn from a specific distribution rather than from deterministic data~\cite{PhysRevLett.106.220503}. Additionally, many problems are ill-conditioned or suffer from the curse of dimensionality, further complicating the reconstruction process.

Various methods have been proposed for solving QST problems. Maximum-likelihood estimation (MLE) methods gained popularity due to their numerical stability and ease of implementation. Iterative MLE methods~\cite{PhysRevA.75.042108} were then introduced, which use iterative computation rather than global optimization, significantly reducing the problem-solving complexity. These methods have become prevalent~\cite{Guo2024, PhysRevA.111.022601,Kawasaki2024}. To incorporate prior knowledge, Bayesian approaches~\cite{PhysRevA.85.052120, Granade_2016, Chapman:22} have been proposed. Gradient descent algorithms have also been developed for QST~\cite{gaikwad2025gradientdescentmethodsfastquantum}. More recently, machine learning techniques have been applied to QST, with methods such as basic neural networks, as well as advanced architectures like Generative Adversarial Networks~\cite{PhysRevLett.127.140502} (GANs) and Variational Autoencoders~\cite{doi:10.1142/S0219749921400050} (VAEs). Many works~\cite{RevModPhys.81.299,PhysRevResearch.4.033220,PhysRevA.108.042430,PRXQuantum.6.010303} have demonstrated the feasibility of these methods.
However, several approaches ignore physical constraints and can output non-physical states. Iterative algorithms have been reported to be slow in convergence~\cite{PhysRevResearch.6.033034}, while neural-network schemes converge faster but depend strongly on the chosen loss function and lack the guarantee of reaching the optimal solution~\cite{WU2024109169}.

Parallel to the above progress, convex optimization methods~\cite{PhysRevApplied.18.044041} have been explored. Convex optimization~\cite{boyd2004convex} refers to optimization problems that are subject to convex constraints and convex objective functions. If a problem can be formulated as a convex optimization problem, it can be solved numerically, and the optimal solution can be obtained. Fortunately, most state tomography problems, including bosonic state QST problems, are convex optimization problems, which means that can be effectively solved within this framework. The pioneering work of Ref.~\cite{PhysRevApplied.18.044041} already demonstrated the practical value of casting bosonic state QST as a convex program, providing an important proof-of-principle. Nevertheless, its exploration was largely confined to single-mode states, moderate Hilbert space dimensions, and did not consider flying modes, which are propagating quantum field modes that carry quantum information through a transmission line rather than being stored in a localized resonator\cite{kimble2008quantum}.

In this work, we efficiently solve high-complexity state reconstruction problems, with a sample-based maximum-likelihood estimation (MLE) method, and three techniques under the convex optimization framework. In detail, a sample-based MLE method is developed for flying-mode tomography. We also investigate convex optimization as a practical framework for bosonic state QST. Within this framework, we introduce three techniques that improve efficiency and scalability: fast evaluation of displacement operators, systematic Hilbert space truncation strategy, and stochastic convex optimization. Together, the above three techniques jointly mitigate the principal shortcomings of existing approaches: high computational overhead, limited scalability, and inefficiency in achieving optimality, thereby providing a more robust and efficient framework for bosonic state QST. In addition, numerical simulations confirm that our approach accurately reconstructs demanding states. It efficiently solves flying four-mode problems with a five-level truncation per mode and even eleven-mode problems with a two-level truncation per mode.

This paper is organized as follows. Section \ref{sec:intro} provides the introduction. In Section \ref{sec:QUANTUM STATE TOMOGRAPHY}, we begin with an overview of quantum state tomography and formally define the optimization problems. We discuss flying mode tomography as a special case of quantum state tomography. We then present three main methods for efficiently addressing these problems: efficient displacement operator computation, Hilbert space truncation, and stochastic convex optimization. Section \ref{sec:COMMON APPLICATIONS} presents details of our method and demonstrates its effectiveness through numerical simulation with generated data. Finally, Section \ref{sec:Summary} offers conclusions and a summary of our work.

\section{\label{sec:QUANTUM STATE TOMOGRAPHY}EFFICIENT QUANTUM STATE TOMOGRAPHY}

\subsection{Quantum State Tomography for Bosonic States}

We formally define a QST problem as a constrained optimization problem:
\begin{equation}
    \begin{array}{cl}
\min & d\left(M_{\rho_{\rm {sys}}}, M_\rho\right), \\
\text { s.t. } & \rho \succeq 0, \\
& \operatorname{Tr} \rho=1.
\end{array}
\label{eq:1}
\end{equation}
$M_\rho$ denotes the mapping from the density matrix $\rho$ to an empirical or theoretical data representation. $\rho_{\rm{sys}}$ is the real state of the system that is being measured, and $M_{\rho_{\rm {sys}}}$ corresponds to the experimental measurement value. 
$M_\rho$ can be any measurement result that associates to the density matrix $\rho$ with the Born rule~\cite{Hall2013} like
\begin{equation}
M_\rho= \operatorname{Tr}(\rho \Pi).
\end{equation}
$\Pi$ denotes the measurement operator. 
In homodyne detection, $\Pi$ is usually the quadrature operator $X(\phi)$, which can be defined as  $X(\phi) \equiv \frac{1}{2}\left(a e^{-i\phi} + a^\dagger e^{i\phi}\right)$, where $a $ is the bosonic field annihilation operator. The Wigner and Husimi-Q distributions, as quasiprobability distributions\cite{carmichael2000statistical, carmichael2008statistical}, provide phase-space representations of the quantum state. In phase-space sampling, $\Pi$ corresponds to measurements of the Wigner or Husimi-Q distributions~\cite{harriman1993husimi,PhysRev.40.749,doi:10.1142/S2251158X12000069}. Phase-space measurements are the main measurement families considered in this work. Although the generalized Q-function can also be handled in this framework, we mainly focus on the Q function and Wigner function.

In Eq. (\ref{eq:1}), $d$ represents a two-parameter metric, which can be a distance metric or a divergence metric~\cite{renyi1961measures}. The most commonly used form of metric $d$ is the $l_2$ distance, which is $d(a, b) = \|a - b\|_2^2$.

Here we take the $M_\rho$ as the Q function:
\begin{equation}
Q_\rho(\alpha) = \operatorname{Tr}(\rho \Pi(\alpha)).
\end{equation}
$\Pi(\alpha)$ reduce to $\Pi(\alpha)=\frac{1}{\pi}|\alpha\rangle\langle\alpha|$, and case with noise
\begin{equation}
\label{eq:rho_h}
    \Pi(\alpha)=\frac{1}{\pi} D(\alpha)\rho_hD^\dagger(\alpha),
\end{equation}
where $D(\alpha)$ is the displacement operator, which is defined as $D(\alpha) = \exp(\alpha a^\dagger - \alpha^* a)$, $\rho_h$ is the quantum state of the pure noise (without signals).

To make it easier to calculate, we can first put them in finite-dimensional space and then vectorize the $\rho$ using the vectorize operator $\operatorname{vec}$ (column-major, after truncating to finite dimension). Therefore, the Q function can be written as
\begin{equation}
Q_\rho(\alpha_i)=\operatorname{vec}\left(\Pi(\alpha_i)^T\right)^T\operatorname{vec}(\rho),
\end{equation}
and the measurement results as $Q_{\rho_{\rm{sys}}}(\alpha_i)$.
Then we stack a set of measurement data points to form a matrix representation as
\begin{equation}
\label{eq:Anb}
    A = \begin{pmatrix} 
\operatorname{vec} \left( \Pi(\alpha_1)^T \right)^T \\
\operatorname{vec} \left( \Pi(\alpha_2)^T \right)^T \\
\vdots \\
\operatorname{vec} \left( \Pi(\alpha_B)^T \right)^T
\end{pmatrix}, \quad b = \begin{pmatrix} 
Q_{\rho_{\rm{sys}}}(\alpha_1) \\
Q_{\rho_{\rm{sys}}}(\alpha_2) \\
\vdots \\
Q_{\rho_{\rm{sys}}}(\alpha_B)
\end{pmatrix}.
\end{equation}
The problem can be written as 
\begin{equation}
\label{eq:amb}
\begin{array}{cl}
\min & \|A \operatorname{vec}(\rho) - b\|_2^2, \\
\text { s.t. } & \rho \succeq 0, \\
& \operatorname{Tr} \rho=1.
\end{array}
\end{equation}
This method is suitable for problems with a small number of modes, and the Q function can be directly measured~\cite{Kirchmair2013} or counted by the histogram method~\cite{onoe2023direct}.

Wigner function can be written as 
\begin{equation}
W_\rho(\alpha)=\frac{2}{\pi}\operatorname{Tr}\left(\rho D(\alpha)PD^\dagger(\alpha)\right),
\end{equation}
where $P=e^{i\pi a^\dagger a}$ is the parity operator~\cite{PhysRevA.60.674}. State tomography with Wigner function measurements can also be written as the same form of problem (\ref{eq:amb}) with $\Pi(\alpha) = \frac{2}{\pi}D(\alpha)PD^\dagger(\alpha)$ .

If we take the metric $d$ as Kullback-Leibler divergence~\cite{10.1214/aop/1176996454} and $M$ as the $Q$ function (a common case in heterodyne detection with linear amplifiers for flying photon measurements), the optimization objective function can be written as
\begin{equation}
d\left(M(\rho_{\rm{sys}}),M(\rho)\right)=\int Q_{\rho_{\rm{sys}}}(\alpha)\log\frac{Q_{\rho_{\rm{sys}}}(\alpha)}{Q_{\rho}(\alpha)}d^2\alpha,
\end{equation}
where $Q_{\rho_{\rm{sys}}}$ represents real distribution, i.e. an unknown but definite function, so $\operatorname*{min}\int Q_{\rho_{\rm{sys}}}(\alpha)\log\frac{Q_{\rho_{\rm{sys}}}(\alpha)}{Q_{\rho}(\alpha)}d^2\alpha$  is equivalent to $\min-\int Q_{\rho_{\rm{sys}}}(\alpha)\log Q_\rho(\alpha)d^2\alpha$. When the samples $\alpha$ follow the distribution $Q_{\rho_{\rm{sys}}}$, for example in quadrature detection with linear amplifier, the problem can be expressed as
\begin{equation}
\min_{\rho\in \mathcal{K}}\underset{\alpha\sim Q_{\rho_{\rm{sys}}}}{\operatorname*{\operatorname*{E}}} \left[-\log Q_\rho(\alpha)\right].
\end{equation}
Approximating expectation by average, we have
\begin{equation}
\label{eq:Mst}
\min_{\rho\in \mathcal{K}}\frac{1}{B}\sum_{n=0}^B-\log Q_\rho(\alpha_n),
\end{equation}
where $B$ is the size of the dataset and $\mathcal{K}$ is the feasible region of $\rho$. This form can also be treated as the maximum likelihood estimation for continuous variables. We refer to this approach as the sample-based MLE method. Unlike histogram-based methods, it does not rely on discretizing the measurement outcomes, which enables it to efficiently handle multi-mode problems. This formulation becomes challenging to solve as the problem scale increases, as discussed in the Appendix \ref{adx:limitation}. The following section outlines how to employ stochastic convex optimization~\cite{shalev2009stochastic, bubeck2015convex, xu2017stochastic} to handle this problem.

\subsection{Efficient Displacement Operator Computation}
In many experiments employing quadrature detection or parity measurements  (using the $Q$ function or the Wigner function) for tomography, the efficient calculation of the displacement operator $D(\alpha)$ is crucial.
However, direct calculation of $D(\alpha)$  is computationally expensive because it involves numerous matrix exponential calculations. Here we introduce a method for numerically efficiently calculating the displacement operator $D(\alpha)$ denoted EDOC (Efficient Displacement Operator Computation). Moreover, mini-batching is incorporated into this method to further accelerate the computation. 

The main idea is to accelerate the calculation of $D(\alpha)$ by taking advantage of the shared eigenvector property of the tridiagonal matrices~\cite{KREER199465} (in finite-dimensional space). For example, $\alpha a^\dagger - \alpha^* a$ and $a+a^\dagger$ are similar for any $|\alpha| = 1 $. Details can be found in the Appendix \ref{adx:EDOC}.

Suppose $\alpha = r e^{i\theta}$, and matrices are truncated into dimension $k$, then the final result is
\begin{equation}\label{eq:efficient_displacement_operator}\begin{gathered}
D(\alpha)\approx
K(\theta)^{-1}V\exp(-ir\Lambda)V^\dagger K(\theta).
\end{gathered}\end{equation}
Both $K$ and $\Lambda$ are diagonal matrices, and 
\begin{equation}
K(\theta) = \operatorname{diag}\Big(1, -i\exp(-i\theta),\dots,(-i\exp(-i\theta))^{k-1}\Big).
\end{equation}
Here, $V$ and  $\Lambda$ correspond to eigenvectors and eigenvalues of $a+a^\dagger$ in finite dimension $k$, respectively.  

Typically, the matrix exponential is computed using the Padé approximation~\cite{baker1961pade}. In contrast, this method only requires the reciprocal and exponential of a vector, involves fewer matrix multiplications (one matrix multiplication for each $\alpha$), and does not require matrix inversion. 
In theory, for a single mode that truncated to dimension $N$ with total $m$ modes. EDOC requires only one matrix multiplication, and the computational complexity is about $M(N^m)$, $M(\cdot)$ is for the time cost of single matrix multiplication. The Padé approximation typically involves about fifteen~\cite{moler2003nineteen}, thus have a complexity of $15M(N^m)$. However, because Padé implementations often include more complex control flow, the actual performance may vary in practice. EDOC also incurs almost no additional memory usage, whereas Padé approximation can require up to $8N^2$ extra memory in the worst case for a single computation~\cite{al2010new}.

\begin{figure}
    \centering
   
    \includegraphics[width=1.0\linewidth]{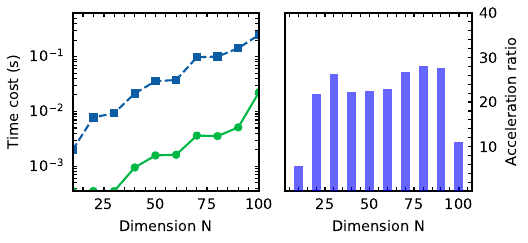}
    \caption{Comparison of time cost and speed acceleration between the proposed method EDOC (green line with circle marker) and the Padé approximation method (blue line with square marker) across different dimensions $N$. The time cost is shown as a line plot on the left y-axis, while acceleration ratio is represented as a bar chart on the right y-axis. The acceleration ratio indicates the ratio of the time cost of the Padé approximation method to that of the EDOC.}
    \label{fig:speed_acceleration}
\end{figure}

We present a speed acceleration test in Fig. \ref{fig:speed_acceleration}. In this test, we randomly generate $\alpha$ and calculate them in mini-batches. The batch size is set to 512, which is a commonly used size in gradient descent methods. The lines represent the time cost for calculating one batch of two methods. Under this setup, we achieve an average speed acceleration of approximately 20.

\subsection{Hilbert Space Truncation}
As is well known, numerical calculations can only handle finite-dimensional problems. In the context of QST, operators in the infinite-dimensional Hilbert space are truncated to a finite-dimensional space. For discrete-variable QST problems (like Fock state tomography with Pauli measurements), Hilbert space truncation may not introduce precision loss, as a finite space can fully capture all the information of a discrete-variable measurement basis (such as the Fock basis). However, in most bosonic state QST problems, there is always a need to balance the space dimension with the resulting precision loss. Here we introduce a strategy to systematically handle finite-dimensional truncation. This strategy is denoted HST (Hilbert Space Truncation).

We take coherent state representation in finite-dimensional space as an example. There are two construction methods, operator method and analytic method. 
Analytic method calculates coherent state $|\alpha\rangle$ by
\begin{equation}
|\alpha\rangle=e^{-\frac{|\alpha|^2}{2}}\sum_{n=0}^\infty\frac{\alpha^n}{\sqrt{n!}}|n\rangle.
\label{eq:analytical}
\end{equation}
While the operator method calculates coherent state $|\alpha\rangle$ by
\begin{equation}
|\alpha\rangle=D(\alpha)\left|0\right\rangle,
\end{equation}
which is more widely used than calculated by Eq. (\ref{eq:analytical}). Thus, it is necessary to increase the number of truncated dimensions to ensure that the values generated by the operator method are closer to the analytical results in lower dimensions.

The truncation operator can be represented by a finite set of orthogonal bases. Defining the orthogonal projection operator $P_k:H\to H_k$, where 
\begin{equation}
P_k=\sum_{n=0}^k|n\rangle\left\langle n\right|.
\end{equation}
For the projection onto an operator $O$, we use the following notation for simplification by defining
\begin{equation}
    [O]_k = P_k O P_k.
\end{equation}

For calculating $\operatorname{Tr}(\rho D(\alpha)\rho_hD^\dagger(\alpha))$ in phase space, there are two steps of projection that need to be used. The first step is to project the operator from infinite dimensions onto a finite dimension without significant loss of precision. The second step is to remove redundant dimensions when calculating the trace of matrix multiplication.
Suppose we constrain $\rho$ within a small photon number space that has dimension $k_1$,  i.e. dimension of the density matrix under test, and we need to calculate $D(\alpha)$ and $\rho_h$ in a higher dimension $k_2$ to maintain precision, i.e. dimension of the measurement operator, where $k_1 \ll k_2$. $\rho_h$ is the pure noise density matrix that has the same definition as it in Eq. (\ref{eq:rho_h}). 

In numerical simulation, we need to align the dimensions of the components when performing matrix multiplication. We can apply $P_{k_1}$ to the second part. The correctness of this step is ensured by the properties of the trace operator. The final result can be calculated as
\begin{equation}
\label{truncated}
    \begin{aligned}
    &\operatorname{Tr}\left(\left[[\rho]_{k_1}\right]_{k_2} [D(\alpha)]_{k_2}[\rho_h]_{k_2}[D^\dagger(\alpha)]_{k_2}\right) \\
    &= \operatorname{Tr}\left([\rho]_{k_1} \left[[D(\alpha)]_{k_2}[\rho_h]_{k_2}[D^\dagger(\alpha)]_{k_2}\right]_{k_1}\right).
\end{aligned}
\end{equation}

Because the latter form requires a $k_1 \times k_1$ matrix multiplication rather than $k_2 \times k_2$, it is computationally more efficient and is therefore preferred in practice. For multi-dimensional cases, such as the two-dimensional case, the operator can be defined as
\begin{equation}
P_{k, m} = P_k \otimes P_m. \end{equation}
As an example of a multi-mode system, consider the two-mode system with the following $Q$ function (Different subscripts represent different subsystems)
\begin{equation}
Q(\alpha,\beta)=\frac{1}{\pi^2}\operatorname{Tr}\left(\rho(D_1(\alpha)\otimes D_2(\beta))\rho_h(D_1^\dagger(\alpha)\otimes D_2^\dagger(\beta))\right).
\end{equation}
If we apply a projection operator to reduce the dimensions (for simplicity, we omit the projection operators applied to $D(\alpha)$ and $\rho_h$), the expression of the projection term becomes 
\begin{equation}
\label{eq:proj}
\Big[([D_1(\alpha)]_{d_1}\otimes [D_2(\beta)]_{d_2})\rho_h([D_1^\dagger(\alpha)]_{d_1}\otimes [D_2^\dagger(\beta)]_{d_2})\Big]_{k,m}.
\end{equation}
For the situation that noises of two modes are independent, and their subsystems have truncated dimensions of $d_1$ and $d_2$: 
\begin{equation}\rho_h=[\rho_{h,1}]_{d_1}\otimes[\rho_{h,2}]_{d_2},\end{equation}
where Eq. (\ref{eq:proj}) can be written as
\begin{equation} \label{eq:proj_optimize}
\begin{split}
 &\left[ [D_1(\alpha)]_{d_1}[\rho_{h,1}]_{d_1}[D_1^\dagger(\alpha)]_{d_1}\right]_k  \\
 &\quad \otimes 
 \left[ [D_2(\beta)]_{d_2}[\rho_{h,2}]_{d_2}[D_2^\dagger(\beta)]_{d_2}\right]_m.
\end{split}
\end{equation}
We simply calculate the two modes separately, truncate them individually, and then apply the Kronecker product. Using Eq. (\ref{eq:proj}) requires computing a matrix multiplication of dimension $d_1 \times d_2$, while Eq. (\ref{eq:proj_optimize}) only involves matrix multiplication of dimensions $d_1$ and $d_2$ individually. 
For a single mode of dimension $N$ and $m$ total modes, Eq. (\ref{eq:proj}) entails a matrix multiplication of time cost $M(N^m)$ ($M(\cdot)$ for cost of single matrix multiplication), while Eq. (\ref{eq:proj_optimize}) only involves $m$ matrix multiplications of time cost $M(N)$. 
Since the computational complexity of matrix multiplication scales cubically with respect to the matrix dimension, the latter formulation leads to a substantial reduction in computational cost. The memory cost of both methods is considered as similar.

Providing a tight bound on the error is non-trivial, as Eq. (\ref{truncated}) involves two truncation steps. The first truncation arises when evaluating the measurement operator $[D(\alpha)]_{k_2}[\rho_h]_{k_2}[D^\dagger(\alpha)]_{k_2}$ in dimension $k_2$, which introduces an initial source of error. The second truncation further restricts this operator to a subspace of dimension $k_1$. This step actually mitigates the overall error, since only a portion of the initial truncation error propagates.

Below, we discuss how to select $k_2$ to provide a suitable upper bound on the total error.
Suppose a state $\ket \psi$ has its highest population in photon number $n$, the displaced state $D(\alpha)\ket{\psi}$ has the photon number distribution $p_n(l) :=|\bra{l}D(\alpha)\ket{n}|^2$ ($l\geqslant n)$ as~\cite{PhysRevA.41.2645} 
\begin{equation}
    p_{n}(l)=\frac{n!}{l!} |\alpha|^{2(l-n)} e^{-|\alpha|^{2}}\left[\mathcal{L}_{n}^{(l-n)}\left(|\alpha|^{2}\right)\right]^{2},
\end{equation}
where $L_n^{(k)}(\cdot)$ is the generalized Laguerre polynomials.
If we require the photon-number truncation error to be smaller than a prescribed threshold $\epsilon$, we need to choose a truncation cutoff $N$ in which $\sum^\infty_{l=N} p_n(l) < \epsilon$.

There is an empirical rule for choose the truncation. The average photon number of displaced state is $n + |\alpha|^2 $ and the photon number variance is $(2 n +1)|\alpha|^2$. Therefore, an empirical lower bound for the required truncated dimension can be expressed as $n + |\alpha|^2 + k \sqrt{2 n +1}|\alpha|$ where $\alpha$ is the maximum displacement value. $k \geqslant 4$ will be a safe value. Ref.~\cite{provaznik2022taming} gives a detailed discussion of the truncation error in continuous variable systems. If we know the target state can be described in low-dimensional space, then we can choose the above dimension (such as $n + |\alpha|^2 + 4 \sqrt{2 n +1}|\alpha|$) to calculate the measurement operator and then project it into the same low dimension as the target state.  Note that this bound is conservative; in practical experiments, the required truncated dimension is often considerably lower. 


As we always use thermal state as the noise state, we will use displaced thermal state as the measurement operators. Suppose the thermal noise state is defined as $\rho_{\mathrm{th}}$ with average photon number $\overline{n}$. The displaced thermal state is defined as $\rho_{\mathrm{dts}} = D(\alpha)\rho_{\mathrm{th}}D(-\alpha)$. Then, the closed-form~\cite{marian1993squeezed} of $\rho_{\mathrm{dts}}$ can be written as 
\begin{equation}
\begin{aligned}
    \langle m| \rho_{\mathrm{dts}}|l\rangle=\left(\frac{m!}{l!}\right)^{1 / 2} \exp \left(-\frac{|\alpha|^{2}}{\bar{n}+1}\right) \alpha^{l-m} \\
    \frac{\bar{n}^{m}}{(\bar{n}+1)^{l+1}} L_{l}^{(l-m)}\left(-\frac{|\alpha|^{2}}{\bar{n}(\bar{n}+1)}\right),
\end{aligned}
\end{equation}
For situations when $\overline{n}$ or $|\alpha|$ is large, using analytical form will be more efficient.

\subsection{Stochastic Convex Optimization}
In the case of a large number of modes, the density matrix dimension will have a very large dimension (e.g., exceeding 100). When the number of samples $M$ is also large, calculating all the values in one time will require substantial computer memory. There is no convergence guarantee of the optimal solution if we use stochastic gradient descent, such as Riemannian gradient descent~\cite{absil2009optimization, sato2021riemannian}. 

Also, it becomes more difficult to calculate the value of $Q(\alpha)$ based on the histogram method. For example, if the quadratures of each mode are divided into 32 grids, a single mode needs $32 \times 32 = 1024$ grids. For two modes, the number of grids increases to $32 \times 32 \times 32 \times 32 = 1048576$. For four modes, the number of grids increases to the level of $1\times 10^{12}$, which is not a rational size that can be effectively calculated. 
Therefore, we adopt stochastic convex optimization, which guarantees convergence to the optimal solution without the need of storing all the raw data~\cite{shalev2009stochastic}.The procedure can be executed online: acquire a batch of data, process it, discard it, and repeat.

The problem of stochastic convex optimization is defined as
\begin{equation}\underset{\rho\in\mathcal{K}}{\operatorname*{\operatorname*{argmin}}}F(\rho),\end{equation}
where $F(\rho)=\operatorname{E}_Z[f(\rho;Z)]$, and for all $\forall z_i \in Z$, $f(\rho; z_i)$ is a convex function with respect to $\rho$. $E$ is the expectation operator. $Z$ represents the dataset. $\mathcal{K}$ is the domain of $\rho$.
Learning from sampled data $z_i$ is equivalent to minimizing the empirical risk~\cite{vapnik1991principles}
\begin{equation}F(\rho)=\operatorname{E}_Z[f(\rho;Z)]\approx\frac{1}{n}\sum_{i=1}^nf\left(\rho;z_i\right).\end{equation}

In our problem, i.e., Eq. (\ref{eq:Mst}), if the value of $B$ is very large, we can use the stochastic gradient descent method, by using a mini-batch each time. In detail, whole train dataset with size $B$ is evenly split into $n$ smaller subsets, each called a mini-batch. Each mini-batch is a random collection of samples used to update the parameters being learned. We can use the average of multiple mini-batches to approximate $\operatorname{\mathrm{E}}_{\alpha\sim Q_{\rho_{\rm{sys}}}}-\operatorname{log}Q_{\rho}(\alpha)$ as 
\begin{equation}
\underset{\alpha\sim Q_{\rho_{\rm{sys}}}}{\operatorname*{\operatorname{E}}}-\log Q_\rho(\alpha)\approx\frac{1}{n|B_i|}\sum_{i=0}^n\sum_{\alpha_j\in B_i}-\log Q_\rho(\alpha_j),
\end{equation}
where $B_i$ is a mini-batch sampled from the whole dataset with size $|B_i|$. 

Thus, the problem can be written as 
\begin{equation}
\begin{array}{cl}
\min & \frac{1}{n}\sum_{i=0}^n\|A_i \operatorname{vec}(\rho) - b_i\|_2^2, \\
\text { s.t. } & \rho \succeq 0, \\
& \operatorname{Tr} \rho=1,
\end{array}
\end{equation}
for sampling rows from matrix $A$ and $b$ in Eq. (\ref{eq:amb}), 
\begin{equation}
    \begin{array}{cl}
\min &\frac{1}{n|B_i|}\sum_{i=0}^n\sum_{\alpha_j\in B_i}-\log Q_\rho(\alpha_j), \\
\text { s.t. } & \rho \succeq 0, \\
& \operatorname{Tr} \rho=1,
\end{array}
\end{equation}
for maximum likelihood method.
Furthermore, we can apply accelerated stochastic convex optimization methods to achieve significant acceleration.

\subsection{Solving Convex Optimization}
There are several methods for solving convex optimization problems, such as the simplex method, interior-point methods, and projection-based approaches. However, these methods do not always perform well when applied to QST problems. Additionally, software packages like CVXpy~\cite{diamond2016cvxpy} are commonly used to solve standard convex optimization problems.

In the context of bosonic state QST, the matrix involved in optimization problems (as shown in problem (\ref{eq:amb})) tends to be relatively dense, especially compared to discrete-variable problems. Since most solvers in CVX are designed to handle sparse matrices, they are not well suited for the complexities of higher-dimensional QST problems. 

For higher-dimensional problems, the projected gradient descent (PGD) method~\cite{Gonçalves03032016}, which can handle dense matrices, is a more appropriate choice. Its stochastic counterpart is denoted stochastic projected gradient descent (SPGD). This method is a specific case of the proximal gradient method, which ensures optimal solutions and provides strong convergence guarantees. 

The projection operator onto set $\mathcal{K}$ is defined as
\begin{equation}
\label{eq:proj_op}
     \operatorname{proj}_{\mathcal{K}}(\rho)=\underset{z\in \mathcal{K}}{\operatorname{argmin}}\|z-\rho\|^{2}.
\end{equation}
In our problem, $\mathcal{K}$ is defined as $\mathcal{K} = \{\rho | \rho\succeq0, \operatorname{Tr}\rho = 1\}$. There exists a closed-form solution~\cite{Gonçalves03032016, Bolduc2017} when the matrix norm used in Eq. (\ref{eq:proj_op}) is the Frobenius norm.
Suppose the eigenvalue decomposition of $\rho$ is given by
\begin{equation}
    \rho = V^\dagger \operatorname{diag}(v) V, 
\end{equation}
where $v$ is the vector of eigenvalues of $\rho$. Then, the projection of $\rho$ onto $\mathcal{K}$ is given by
\begin{equation}
    \operatorname{proj}_{\mathcal{K}}(\rho) = V^\dagger \operatorname{diag}(\operatorname{proj_\Delta}(v)) V,
\end{equation}
where $\operatorname{proj_\Delta}$ denotes the simplex projection operator applied to the vector. A detailed discussion of this method, including its advantages and limitations, is provided in Appendix \ref{adx:pgd}.
The stochastic version for low-memory calculation is detailed in Appendix \ref{adx:stochastic}.


\section{\label{sec:COMMON APPLICATIONS}COMMON APPLICATIONS}
We evaluate our method on both simple cases involving single-mode and two-mode bosonic state QST problems, as well as more complex cases with more than four modes. For the single-mode cases, we utilize CVXpy (using SCS~\cite{odonoghue:21,aa2020} as the solver) to solve the problems for simplicity. In contrast, for cases with more modes, we employ the PGD and SPGD~\cite{xu2017stochastic} solver to solve the problems. Stochastic convex optimization techniques are applied to address the challenges posed by the high dimensionality of the more-than-four-mode problems, which would be difficult to implement without stochastic methods due to memory and computational limitations. The sample-based MLE approach is used to handle flying-mode problems.

\subsection{single mode case}
\begin{figure}
    \centering
    \includegraphics[width=1\linewidth]{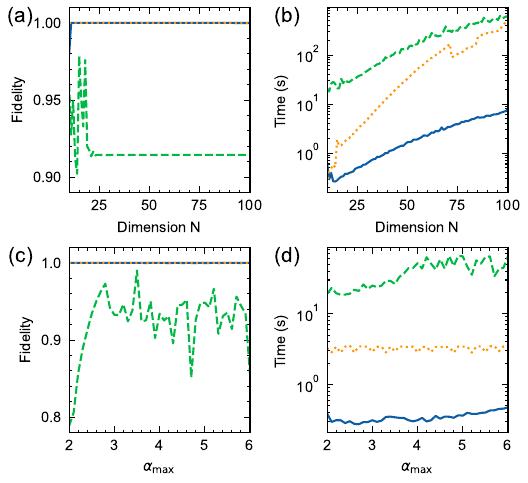}
    \caption{Single-flying‐mode quantum‐state reconstruction with histogram method. Figures show the performance of three algorithms as a function of Hilbert‐space dimension \(N\) and phase‐space limit \(\alpha_{\mathrm{max}}\). Blue lines are for EDOC+HST+SCS, orange dotted lines are for baseline convex optimization method and green dashed lines are for MLE method. (a) Reconstruction fidelity versus \(N\). (b) Computational runtime versus \(N\) (logarithmic scale). (c) Reconstruction fidelity versus \(\alpha_{\mathrm{max}}\). (d) Computational runtime versus \(\alpha_{\mathrm{max}}\) (logarithmic scale).}
    \label{fig:1}
\end{figure}

The single-mode problem is the simplest and the basis for higher-dimensional problems. Here we reconstruct the single-mode density matrix from noiseless heterodyne data that is generated with the Markov chain Monte Carlo (MCMC) method~\cite{10.1214/10-STS351}. 
We set the test state $|\phi\rangle$ to a Cat state with $\alpha=2$, defined as
\begin{equation}
    |\phi\rangle \propto \ket{\alpha}+\ket{-\alpha}.
\end{equation}
We change the dimension of the density matrix under test $N$ and set the dimension of the measurement operator to $\lceil1.5N\rceil$ . We first set the grid size to $20 \times 20$ and the phase space limit $\alpha_{\rm max} = 4$ and measured the overall runtime for solving the problem in Figs. \ref{fig:1}(a) and (b). The results indicate that both the baseline convex optimization method (in Ref.~\cite{PhysRevApplied.18.044041}) and the proposed method achieve a reconstruction fidelity exceeding 99.9\%, while the MLE method fails to reach such a high level of fidelity. Moreover, as the dimension $N$ increases, our proposed method demonstrates a clear advantage in terms of problem-solving time—solving the problem in under 10 seconds even when $N=100$. Overall, our results demonstrate that the proposed method is approximately ten times faster without compromising reconstruction fidelity.

Figures \ref{fig:1}(c) and (d) show the fidelity and computation time as functions of the phase space limit $\alpha_{\rm max}$ with grid size of $15 \times 15$ and $N=32$. Overall, the phase space limit does not significantly impact the final fidelity or the computation time.

\subsection{two modes case}
\begin{figure}
    \centering
    \includegraphics[width=1\linewidth]{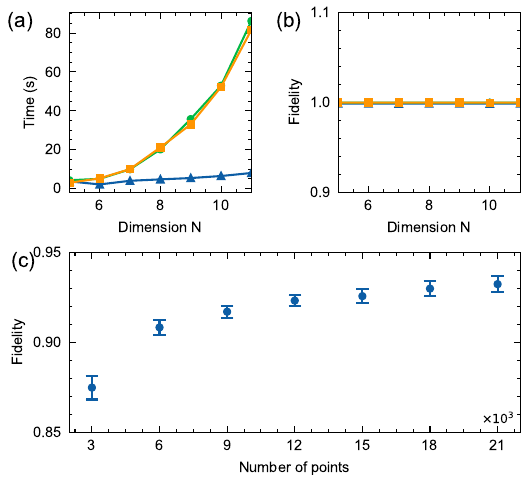}
    \caption{Two-stationary‐mode state reconstruction from Wigner functions. (a) Computational runtime versus single mode dimension $N$ (total dimension is $N^2$). Blue lines with triangle markers represent EDOC+HST+PGD, while orange with square markers and green lines with circle markers correspond to EDOC+HST+SCS and the baseline convex optimization method, respectively. (b) Fidelity versus single mode dimension $N$ (total dimension is $N^2$). Line colors correspond to those in panel (a).  (c) Fidelity versus the number of data points with $N=6$. The batch size is fixed at 3000, and the number of data points is varied by changing the number of batches used in the optimization. Error bars indicate the standard deviation of fidelity over different random data initializations.}
    \label{fig:4}
\end{figure}

Two-mode system is more complicated. In this task, we reconstruct the target density matrix from Wigner function data. 
We use a two-mode logical binomial Bell state as the target state, which is widely used in quantum communication and quantum error correction. The state here is a discrete-variable state, but we use continuous-variable measurement bases which are Wigner function measurement bases.
The two-mode logical Bell state is defined as 
\begin{equation}
    \ket{\phi} =\frac{1}{\sqrt{2}} (\ket{0_{\rm L}1_{\rm L}}+ \ket{1_{\rm L}0_{\rm L}}),
\end{equation}
with 
\begin{equation}
    \ket{0_{\rm L}} = \frac{\ket{0}+\ket{4}}{\sqrt 2}, \ \ket{1_{\rm L}} = \ket{2}.
\end{equation}
We change the dimension of the density matrix under test $N$ and set the dimension of the measurement operator to 24, batch size to 3000, start learning rate to 100, and maximum training steps to 2000.
We show the fidelity and computation time of this setup with the single mode dimension $N$ (the total computational dimension is $N ^2$) in Fig. \ref{fig:4} (a) and (b). In our setup, the Hilbert space dimension does not have an influence on the final fidelity, which remains at 1.  Moreover, employing PGD as the solver substantially reduces the overall computation time. For the two-mode examples, the cost of constructing the measurement bases is negligible compared with the time required to solve the optimization problem; therefore, improvements in the solver translate directly into reductions in total runtime. This is different from the single mode cases. Although the effective Hilbert-space dimension of the two-mode system is on the order of, for example, 100 , we only compute the single-mode displacement operator in a truncated basis with $N=10$ and obtain the full two-mode operator by taking the Kronecker product of the single-mode operators. Forming the two-mode operator in this way is considerably faster than evaluating the displacement operator directly in a basis of size $N=100$.

In Fig. \ref{fig:4}(c), we illustrate how the number of data points impacts the final fidelity with the effect of noise. We employ SPGD, more specifically, a method called Accelerated Stochastic Subgradient Method: the regularized variant (ASSG-r) as the solver and model the noise as additive Gaussian noise with standard deviation $\sigma=0.1$, which is approximately 25\% of the maximum signal amplitude. The dimension of the density matrix under test is set to 6 and the start learning rate is set to 0.5. The batch size is set to 3000, corresponding to roughly $10~\mathrm{GB}$ of memory. These results show that, using a compute-then-discard strategy, one can increase the number of data points to raise the reconstruction fidelity without increasing the peak memory footprint. 

\subsection{four and more modes case}
Applying CVXpy to solve a four-mode system is not feasible due to its larger Hilbert space dimension with high memory usage. 
Recent works report the technology of a momentum-space reconstruction in low dimension (such as a 4-mode, 2-level subspace) with a tensor-network extension to higher modes \cite{sunada2024efficienttomographymicrowavephotonic,O’Sullivan2025}, but they still rely on an explicit low-dimensional reconstruction and are hard to extend to more-level subspace because the complexity of calculating momentum will increase rapidly.
To address these challenges, we turn to the sample-based MLE method with stochastic convex optimization methods. The main idea of this method is to use likelihood to avoid using a histogram and divide the computation into small batches, and optimize them individually. Additionally, we use ASSG-r to further improve the optimization efficiency. With the help of stochastic convex optimization, we can converge to the optimal solution with nearly 100\% probability even using small batches.

\begin{figure}
    \centering
    \includegraphics[width=1\linewidth]{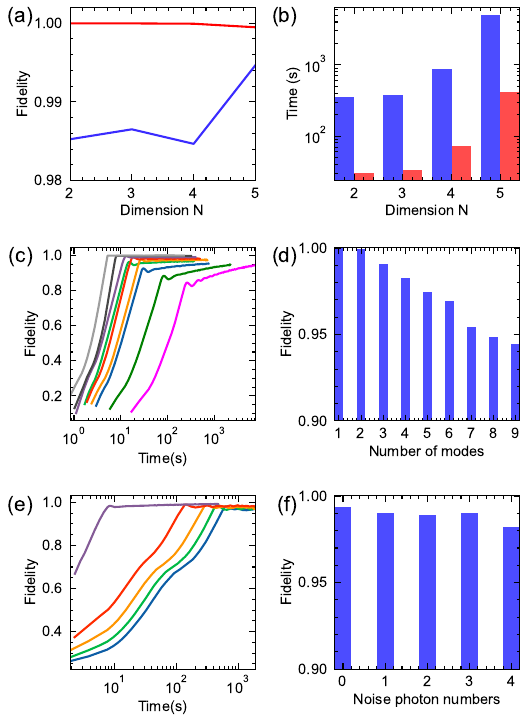}
    \caption{
    Results of flying mode tomography under different setups.
 (a) Final fidelity within 1000 training steps for various single-mode dimensions $N$ (four modes in total); blue: baseline stochastic convex optimization, red: ASSG-r.
 (b) Characteristic decay time of infidelity from panel (a), defined as the training time for the infidelity to drop to $1/e$ of its initial value.
 (c) Reconstruction fidelity versus training time for systems with 1–9 modes (dimension 2 per mode). Color code: gray (1), black (2), dark purple (3), light green (4), red (5), orange (6), blue (7), dark green (8), light purple (9).
 (d) Final fidelity within 1000 steps as a function of the number of modes, corresponding to panel (c).
 (e) Reconstruction fidelity versus training time for different noise photon numbers (purple: 0, red: 1, orange: 2, green: 3, blue: 4).
 (f) Final fidelity within 1000 steps as a function of noise photon numbers, corresponding to panel (e).
    }
    \label{fig:scp_four_mode}
\end{figure}

We use a four-qubit W state as the test state and generate flying mode data using the MCMC method. Each data point is treated as a sample from a $Q$-function. We change the dimension of the density matrix under test $N$ and set the dimension of the measurement operator to 10, batch size to 256, start learning rate to 1000, and the inner iteration number to 200. We first test our method on noiseless data, and the results are shown in Fig. \ref{fig:scp_four_mode}. Figure \ref{fig:scp_four_mode}(a) displays the maximum fidelity achieved within 1000 training steps for a four mode problem, with the accelerated method reaching nearly 100\% fidelity. Figure \ref{fig:scp_four_mode}(b) illustrates how fast the fidelity increases during the training process, demonstrating a significant acceleration with the ASSG-r method. 
We extend the method to handle multiple modes to demonstrate its effectiveness for multi-mode problems. We use a multi-mode W state as the test state and set batch size to 512, inner iteration number to 100, learning rate to 0.005. Figure \ref{fig:scp_four_mode}(c) and (d) display the reconstruction fidelity versus training time for different numbers of modes and the corresponding final fidelity. Notably, this setup enables the reconstruction of a flying 9-mode W state.

When noise is taken into account, the sample-based MLE method remains applicable. Figure \ref{fig:scp_four_mode}(e) displays the fidelity versus training time for different noise photon numbers. In these tests, the Hilbert space of each mode is set to 2, while the computational Hilbert space used for calculating the displacement of the thermal state is set to 40. We observe that as the noise photon number increases, the overall time cost also increases. Notably, at a noise level corresponding to 4 noise photons, a final reconstruction fidelity exceeding 98\% is achieved within 400 seconds—a highly competitive result. Figure \ref{fig:scp_four_mode}(f) presents the final fidelity for various noise photon numbers. All cases achieve a final fidelity exceeding 98\%, and further optimization with a smaller learning rate can enhance these results.

\section{\label{sec:Summary}Summary}

In summary, we have developed convex optimization methods for bosonic state tomography problems. Under the constraints of positivity and normalization, we transform the problem into a convex optimization problem, solvable with guaranteed global convergence. Several practical techniques are introduced to achieve high efficiency. In numerical simulations, we solve the histogram-based problems and two-mode Wigner tomography problems, demonstrating robust performance. For the more challenging four-mode and more modes flying mode tomography problems, we propose a sample-based maximum-likelihood estimation method.  By employing a stochastic proximal gradient descent algorithm (here, SPGD) within this framework, we show that the use of stochastic convex optimization is critical in addressing the high-dimensional challenges posed by these cases, overcoming memory and computational limitations that traditional methods cannot efficiently manage. Comparing with conventional approaches, our methods are more stable, scale better with system size, and avoid common problems such as nonphysical outputs or convergence to local minima. Further, the proposed methods achieve higher reconstruction fidelity and significantly reduce computational cost compared to conventional approaches. Particularly, the framework remains robust under noise and performs well in high-dimensional problems. With these achievements and advantages, the proposed framework can facilitate the development of continuous-variable quantum error correction codes and quantum interconnect with non-Gaussian states, which are essential for fault-tolerant, hardware-efficient quantum architectures.

\begin{acknowledgments}

\red{
This work was supported by the National Natural Science Foundation of China (NSFC) under Grant No. 62433015 and by Laoshan Laboratory (LSKJ202200900). Z.-L. X. was supported by the NSFC under Grant No. 12375025. J. Z. received support from the Leading Scholar Program of Xi’an Jiaotong University, the Innovative Leading Talent Project \enquote{Shuangqian Plan} of Jiangxi Province, and the \enquote{Tianfu Emei Plan}, the Provincial–University–Enterprise Cooperation Talent Special Project. P. S. was supported by the \enquote{Young Talent Support Plan} of Xi’an Jiaotong University. H. I. thanks the support by FDCT of Macau under grants 0179/2023/RIA3 and 0134/2024/AFJ.}

\end{acknowledgments}

\appendix

\section{Efficient Displacement Operator Computation}\label{adx:EDOC}
Here, we give the detailed derivation of equation (\ref{eq:efficient_displacement_operator}):
The definition of the displacement operator can be written as:
\begin{equation}\label{eq:defination_displacement_operator}D(\alpha)=\exp(\alpha a^\dagger-\alpha^*a),\end{equation}
while $\alpha a^\dagger - \alpha^* a$ is a skew-Hermitian matrix. By multiplying $i$ to $\alpha a^\dagger - \alpha^* a$, we can transform it into a tridiagonal Hermitian matrix:

\begin{equation}
\left. iG = i\alpha a^\dagger - i\alpha^* a =
\begin{pmatrix}
0 & b_1 &        &        &        \\
c_1 & 0 & b_2    &        &        \\
    & \ddots & \ddots & \ddots &   \\
    &        & c_{k-2} & 0 & b_{k-1} \\
    &        &        & c_{k-1} & 0
\end{pmatrix}
\right.
\end{equation}
where $b_nc_n =n  |\alpha|^2, b_n = c_n^\dagger, \quad n = 1\dots k-1$.

Additionally, tridiagonal matrices have an important property. For a tridiagonal matrix $T_k$:
\begin{equation}
T_k = \begin{pmatrix}
\alpha_1 & \omega_1 &         &         &           \\
\beta_1  & \alpha_2 & \omega_2&         &           \\
        & \ddots   & \ddots  & \ddots  &           \\
        &          & \beta_{k-2} & \alpha_{k-1} & \omega_{k-1} \\
        &          &         & \beta_{k-1}  & \alpha_k
\end{pmatrix}.
\end{equation}
\begin{widetext}
If $\omega_i \beta_i >0$, we have:
\begin{equation}J_k=K_k^{-1}T_kK_k,\end{equation}
where

\begin{equation}
J_k=
\begin{pmatrix}
\alpha_1 & \sqrt{\omega_1\beta_1} & & & \\
\sqrt{\omega_1\beta_1} & \alpha_2 & \sqrt{\omega_2\beta_2} & & \\
 & \ddots & \ddots & \ddots & \\
 & & \sqrt{\omega_{k-2}\beta_{k-2}} & \alpha_{k-1} & \sqrt{\omega_{k-1}\beta_{k-1}} \\
 & & & \sqrt{\omega_{k-1}\beta_{k-1}} & \alpha_k
\end{pmatrix},
\end{equation}
\begin{equation}K_k=\operatorname{diag}(\delta_1,\ldots,\delta_\mathrm{k}),\delta_1=1,\delta_{j}^2=\frac{\beta_{{j-1}}\cdots\beta_1}{\omega_{{j-1}}\cdots\omega_1},\quad{j=2,\dots, k},\end{equation}
i.e. the matrix $T_k$ is similar to $J_k$, and the invertible matrix $K_k$ is a diagonal matrix. 
According to these properties, we have:

\begin{equation}\left.J_F=r \times \left(
\begin{array}
{ccccc}0 & 1 & & & \\
1 & 0 & \sqrt{2} & & \\
 & \ddots & \ddots & \ddots & \\
 & & \sqrt{k-2} & 0 & \sqrt{k-1} \\
 & & & \sqrt{k-1} & 0
\end{array}\right.\right),\end{equation}

\end{widetext}
\begin{equation}
K(\theta)=\mathrm{diag}{\left(1,-\mathrm{i}\exp(-\mathrm{i}\theta),\ldots,(-\mathrm{i}\exp(-\mathrm{i}\theta))^{\mathrm{k}}\right)},
\end{equation}
where the coefficient of coherent state $\alpha$ is written as $\alpha = r e^{i\theta},\quad r\in \mathcal R, \theta\in[0, 2\pi]$. 

Bring all these results into equation (\ref{eq:defination_displacement_operator}):
\begin{equation} D(\alpha)=\exp(-i \times iG)=K(\theta)^{-1}\exp(J_F)K(\theta),\end{equation}
where $J_F$ is a Hermitian matrix, which can be diagonalized and written as $V\Lambda V^{\dagger}$. $V$ is the eigenvector and $\Lambda$ is the diagonal matrix.

Finally, we can write equation (\ref{eq:defination_displacement_operator}) into equation (\ref{eq:efficient_displacement_operator}).

\section{Convex Optimization Techniques}
\subsection{Limitation of General-Purpose Solvers \label{adx:limitation}}
Generally, the common convex optimization libraries model the problem as cone programming~\cite{alizadeh2003second} (CP), to solve general problems. Obviously, the efficiency of general cone programming methods is worse than specially designed methods for specific problems.

Further, most common solvers like SCS~\cite{ocpb:16}, MOSEK construct a logarithmic target function as an exponential cone constraint. In cases where problems involve the summation of a larger number of logarithmic functions  (such as maximum likelihood estimation), these solvers model the problem as the intersection of an equal number of exponential cone constraints. Unfortunately, multiple exponential cone constraints cannot be handled efficiently, leading to poor performance in these problems.

Another defect is that most common solvers are designed for sparse problems. In CV-QST, the matrix A is often very dense, leading to low efficiency. Thus, for large scale problems, we implement our dense-matrix based solver using Numpy~\cite{2020NumPy-Array} and JAX~\cite{jax2018github}.  Numpy is for CPU-based solver, while JAX is for GPU-based solver.

\subsection{Projected Gradient Descent method\label{adx:pgd}}
Proximal gradient method is used to handle problems that the objective function may not be differentiable. We rewrite the problem (\ref{eq:amb}) into a new form 
\begin{equation}
    \min\quad   g(\rho) + h(\rho),
\end{equation}
where, for example, 
\begin{equation*}
    g(\rho) = \|A \operatorname{vec}(\rho) - b\|_2^2
\end{equation*}
and 
\begin{equation*}
    h(\rho)= I_{\mathcal{K}}(\rho)
\end{equation*}
with $\mathcal{K} = \{\rho | \rho\succeq0, \operatorname{Tr}\rho = 1\}$.
The indicator function $I_\mathcal{K}(\rho)$ is defined as
\begin{equation*}
    I_{\mathcal{K}}(\rho):=\left\{\begin{array}{ll}
0 & \text { if } \rho \in \mathcal{K}, \\
+\infty & \text { if } \rho \notin \mathcal{K} .
\end{array}\right.
\end{equation*}
Proximal operator is defined as
\begin{equation}
    \operatorname{prox}_{h, t}(\rho)=\underset{z}{\operatorname{argmin}} \frac{1}{2 t}\|z-\rho\|^{2}+h(z),
\end{equation}
where $t >0$ is a small scalar factor.

Utilizing the proximal operator in the iterative process ensures both convergence and optimality.
\begin{equation}
\label{eq:pgd}
    \rho^{(k)}=\operatorname{prox}_{h, t_{k}}\left(\rho^{(k-1)}-t_{k} \nabla g\left(\rho^{(k-1)}\right)\right), \   k=1,2,3, \ldots,
\end{equation}
where $k$ is the iteration index and $\nabla$ is the gradient operator. Formula (\ref{eq:pgd}) is called proximal gradient descent.

When $h(\rho) = I_{\mathcal{K}}(\rho)$ is an indicator function, the corresponding proximal operator reduces to a projection onto the set $\mathcal{K}$. Specifically, the proximal operator is defined as
\begin{equation}
    \operatorname{prox}_{I_\mathcal{K}, t}(\rho)=\underset{z\in \mathcal{K}}{\operatorname{argmin}} \frac{1}{2 t}\|z-\rho\|^{2}.
\end{equation}
Since the scaling parameter $t$ does not affect the location of the minimizer, this expression simplifies to the projection operator onto set $\mathcal{K}$ which is defined as
\begin{equation}
     \operatorname{proj}_{\mathcal{K}}(\rho)=\underset{z\in \mathcal{K}}{\operatorname{argmin}}\|z-\rho\|^{2}.
\end{equation}
This forms the foundation of the projected gradient descent (PGD) method.

The use of PGD as a solver offers two key advantages that contribute to accelerated computation. First, by working entirely in the primal domain and discarding the dual formulation, we significantly reduce the problem dimensionality. Second, we employ simplex projection instead of operator splitting methods~\cite{bauschke2017douglas}, thereby decreasing the number of required projections. The trade-off is the loss of dual information, which limits our ability to certify optimality and detect infeasibility directly. Nevertheless, in practice, optimality can often be assessed through gradient norms or objective values, and most physical problems of interest are inherently feasible.

\subsection{Stochastic Gradient Descent and Acceleration\label{adx:stochastic}}
Since projected gradient descent is a special case of proximal gradient descent, here we use the accelerated proximal gradient descent method mentioned in ref \cite{xu2017stochastic}.
For deterministic convex optimization (i.e. opposite to the stochastic convex optimization in the main text), the descent of each step is guaranteed. 
For stochastic convex optimization, we use the following algorithm to solve the problem:

\begin{algorithm}[H]
    \caption{Stochastic subgradient algorithm}\label{alg:SSG}
    \begin{algorithmic}
    \Require
    \State $\mathcal{K}$: the feasible set of parameters, here is $\{\rho| \rho \succeq 0, \operatorname{Tr}\rho=1 \}$
    \State $\rho_0$: initial parameters
    \State $f$: convex objective function of the problem
    \State $\Xi$: dataset of the measurement results
    \State $B$: the mini-batch size
    \State $T$: the number of inner iterations
    \State $E$: the number of outer iterations
    \State $\eta_0$: learning rate at the beginning
    
    \Ensure density matrix $\rho$
    
    \For {$k = 1, \dots, E$ }
        \State $\rho_1^k \gets \rho_{k-1}$
        \State $\eta_k\gets \eta_{k-1}$
        \For{ $\tau=1,\dots, T$}
            \State Sample a batch $\xi_\tau^k$ from $\Xi$ with size $B$
            \State $\rho_{\tau+1}^k \gets \operatorname{proj}_\mathcal{K}[\rho_{\tau}^k - \eta_k \frac{\partial f(\rho_{\tau}^k, \xi_\tau^k)}{\partial\rho_\tau^k}^*]$
        \EndFor
        \State $\rho_k \gets \frac{1}{T}\sum^T_{\tau=1}\rho_\tau^k$
        \State $\eta_k \gets \eta_k /2$
    \EndFor
    \State Output $\rho_k$
\end{algorithmic}
\end{algorithm}

Accelerated stochastic subgradient method is used to achieve faster global convergence than normal stochastic subgradient methods. 
Comparing to traditional stochastic subgradient method (SSG), i.e. the baseline stochastic convex optimization method in main text, the main difference of ASSG-r is the iteration, which can be written as:

\begin{algorithm}[H]
    \caption{Acceleration stochastic subgradient algorithm}\label{alg:ASSG}
    \begin{algorithmic}
    \Require
    \State $\mathcal K$: the feasible set of parameters, here is $\{\rho| \rho \succeq 0, \operatorname{Tr}\rho=1 \}$
    \State $\rho_0$: initial parameters
    \State $f$: convex objective function of the problem
    \State $\Xi$: dataset of the measurement results
    \State $B$: the mini-batch size
    \State $T$: the number of inner iterations
    \State $E$: the number of outer iterations
    \State $\eta_0$: learning rate at the beginning

    \Ensure density matrix $\rho$ 
    \For {$k = 1, \dots, E$ }
        \State $\rho_1^k \gets \rho_{k-1}$
        \State $\eta_k\gets \eta_{k-1}$
        \For{$\tau=1,\dots, T$}
            \State Sample a batch $\xi_\tau^k$ from $\Xi$ with size $B$
            \State $\rho_{\tau+1}^k \gets \operatorname{proj}_\mathcal{K}[(1 -\frac{2}{\tau}) \rho_\tau^k +\frac{2}{\tau}\rho_1^k - \frac{2\eta_k}{\tau}\frac{\partial f(\rho_{\tau}^k, \xi_\tau^k)}{\partial\rho_\tau^k}^*]$
        \EndFor
        \State $\rho_k \gets \frac{1}{T}\sum^T_{\tau=1}\rho_\tau^k$
        \State $\eta_k \gets \eta_k /2$
    \EndFor
    \State Output $\rho_k$
\end{algorithmic}
\end{algorithm}

By applying the above changes, a faster convergence can be achieved compared to the stochastic convex optimization used in the main text. The main idea is solving the problem approximately in a local region around a historical solution iteratively, and the size of this local region gradually decreases as the solution approaches the optimal set.

With all the techniques above, and taking four mode case as an example, the full pipeline can be written as:
\begin{algorithm}[H]
    \caption{Full pipeline for four mode case}\label{alg:full}
    \begin{algorithmic}
    \Require
    \State $\mathcal{K}$: the feasible set of parameters, here is $\{\rho| \rho \succeq 0, \operatorname{Tr}\rho=1 \}$
    \State $\rho_0$: initial parameters
    \State $f$: convex objective function of the problem
    \State $B$: the mini-batch size
    \State $T$: the number of inner iterations
    \State $E$: the number of outer iterations
    \State $\eta_0$: learning rate at the beginning
    
    \Ensure density matrix $\rho$
    \State$\Xi_0$ randomly sample $n$ points in phase space
    \For {$k = 1, \dots, E$ }
        \State $\rho_1^k \gets \rho_{k-1}$
        \State $\eta_k\gets \eta_{k-1}$
        \For{ $\tau=1,\dots, T$} 
            \State Sample a batch $\Xi_{\tau}^k$ from $\Xi_0$ with size $B$
            \State Compute single mode $D(\alpha)$ by EDOC, where $\alpha\in$ $\Xi_{\tau}^k$ 
            \State Assemble $D(\alpha)$ into full measurement operator $A(\alpha)$ by HST 
            \State Get measurement result $\xi_\tau^k$
            \State Compute loss and gradient of target function $f(\rho_{\tau}^k, \xi_\tau^k)$ with measurement operator $A(\alpha)$
            \State Get $\tilde\rho_{\tau}^k$ by updating $\rho_{\tau}^k$ with gradients
            \State Get $\rho_{\tau+1}^k$ by applying projection to $\tilde\rho_{\tau}^k$
        \EndFor
        \State $\rho_k \gets \frac{1}{T}\sum^T_{\tau=1}\rho_\tau^k$
         \State Update the learning rate $\eta_k$ to get $\eta_{k+1}$
    \EndFor
    \State Output $\rho_k$
\end{algorithmic}
\end{algorithm}

\section{Hyperparameter Sensitivity Analysis}

\begin{figure}
    \centering
   
    \includegraphics[width=1.0\linewidth]{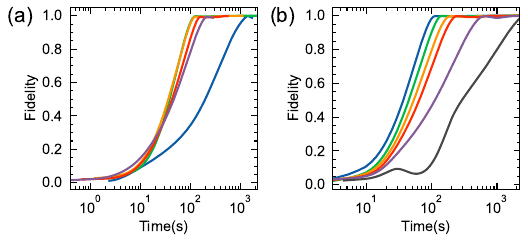}
    \caption{Fidelities versus training time for different mini-batch size $B$ and inner iteration number $T$. (a) Fidelities versus training time for different mini-batch size $B$, $B= 10,25,50,100,200$ for line with color blue, purple, red, orange and green.
    (b) Fidelities versus training time for different inner iteration number $T$, $T=16,32,64,128,256,512$ for line with color black, purple, red, orange, green and  blue.
    }
    \label{fig:s1}
\end{figure}

We first investigate how the mini-batch size $B$ and the inner iteration number $T$ influence the convergence behavior of our method in Fig.~\ref{fig:s1}. By varying these two hyperparameters, we aim to understand their roles in shaping optimization stability, training efficiency, and overall performance. This analysis provides additional insight into selecting appropriate values for reliable and effective training. The simulation parameter setup is the same as Fig.~\ref{fig:scp_four_mode} (a) and the optimizer is ASSG-r.

Figure \ref{fig:s1}(a) illustrates that increasing the batch size generally leads to faster convergence of the solver. However, when the batch size becomes too large, the additional computational cost offsets these gains. A common strategy is to raise the learning rate to further accelerate convergence in large batch size, but this approach also encounters practical limits. This phenomenon is closely related to gradient noise, for which Ref.~\cite{mccandlish2018empirical} provides a detailed discussion. Figure \ref{fig:s1}(b) illustrates how the number of inner iterations $T$ affects learning process. Similar to the effect of batch size $B$, increasing $T$ tends to speed up convergence. In theory, once $T$ exceeds a certain lower bound, the algorithm is guaranteed to converge. In practice, larger $B$ will increase the accuracy of the gradient estimation, thereby accelerate the convergence.

\begin{figure}
    \centering
    \includegraphics[width=1.0\linewidth]{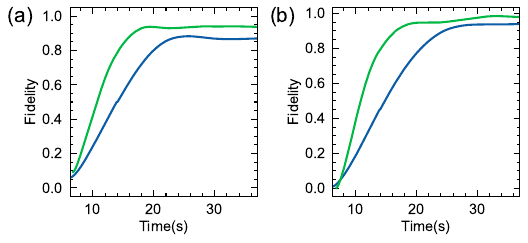}
    \caption{Comparison of uniform sampling and importance sampling. We use two stationary mode Wigner tomography to demonstrate the result. Green lines are for importance sampling and blue lines are for uniform sampling. The learning rate is set to the same value across two subfigures.  (a) shows the comparison with noise level set to $0.1$. (b) shows the comparison without noise.}
    \label{fig:s2}
\end{figure}

We investigated how different sampling strategies affect the learning process, comparing uniform sampling with importance sampling. As shown in Fig.~\ref{fig:s2}(a), importance sampling achieves a higher final fidelity. In the noiseless setting, illustrated in Fig.~\ref{fig:s2}(b), it also converges faster. These observations suggest that importance sampling is a preferable choice for Wigner tomography.

\section{Markov Chain Monte Carlo }
Markov Chain Monte Carlo (MCMC) is used to draw samples from a known probability distribution. MCMC methods are used to study probability distributions that are too complex or too high-dimensional to study with analytic techniques alone. We use a software called PyMC~\cite{Abril-Pla_PyMC_a_modern_2023}.

This method is commonly used to draw samples from a continuous probability distribution and provides an efficient way to generate samples in the phase space of heterodyne detection. In heterodyne detection with a linear amplifier, the measurement result can be denoted as $\alpha$ which follows the distribution $\alpha\sim Q_{}$. 
Therefore, by specifying the form of the $Q$ function, we can directly generate the samples. For example, we could use a specific $Q(\alpha) = \frac{1}{5\pi}\exp{(-|\alpha|^2/5)}$ , or, in the case of a specific density matrix $\rho$, the $Q$ function can be given by $Q(\alpha) = \operatorname{Tr}(\rho |\alpha\rangle\langle\alpha|)$.

\section{Reproducibility and Code Availability }

All numerical experiments were performed on a workstation equipped with a NVIDIA GeForce RTX 4080 GPU and a 13th Gen Intel(R) Core(TM) i9-13900K CPU. The computations used 64-bit floating-point precision and libraries including  CVXpy~\cite{diamond2016cvxpy} , NumPy~\cite{harris2020array}, QuTiP~\cite{JOHANSSON20121760}, dynamiqs~\cite{guilmin2025dynamiqs}, and JAX~\cite{jax2018github} libraries. Details and codes are available on Github~\cite{Littlebaker}.

\nocite{*}

\bibliography{apssamp}

@article{PhysRevLett.105.150401,
  title = {Quantum State Tomography via Compressed Sensing},
  author = {Gross, David and Liu, Yi-Kai and Flammia, Steven T. and Becker, Stephen and Eisert, Jens},
  journal = {Phys. Rev. Lett.},
  volume = {105},
  issue = {15},
  pages = {150401},
  numpages = {4},
  year = {2010},
  month = {Oct},
  publisher = {American Physical Society},
  doi = {10.1103/PhysRevLett.105.150401},
  url = {https://link.aps.org/doi/10.1103/PhysRevLett.105.150401}
}

@article{PhysRevLett.105.250403,
  title = {Permutationally Invariant Quantum Tomography},
  author = {T\'oth, G. and Wieczorek, W. and Gross, D. and Krischek, R. and Schwemmer, C. and Weinfurter, H.},
  journal = {Phys. Rev. Lett.},
  volume = {105},
  issue = {25},
  pages = {250403},
  numpages = {4},
  year = {2010},
  month = {Dec},
  publisher = {American Physical Society},
  doi = {10.1103/PhysRevLett.105.250403},
  url = {https://link.aps.org/doi/10.1103/PhysRevLett.105.250403}
}

@article{PhysRevLett.113.080401,
  title = {Quantum Hamiltonian Identification from Measurement Time Traces},
  author = {Zhang, Jun and Sarovar, Mohan},
  journal = {Phys. Rev. Lett.},
  volume = {113},
  issue = {8},
  pages = {080401},
  numpages = {5},
  year = {2014},
  month = {Aug},
  publisher = {American Physical Society},
  doi = {10.1103/PhysRevLett.113.080401},
  url = {https://link.aps.org/doi/10.1103/PhysRevLett.113.080401}
}

@article{PhysRevLett.108.080502,
  title = {Quantum System Identification},
  author = {Burgarth, Daniel and Yuasa, Kazuya},
  journal = {Phys. Rev. Lett.},
  volume = {108},
  issue = {8},
  pages = {080502},
  numpages = {5},
  year = {2012},
  month = {Feb},
  publisher = {American Physical Society},
  doi = {10.1103/PhysRevLett.108.080502},
  url = {https://link.aps.org/doi/10.1103/PhysRevLett.108.080502}
}

@ARTICLE{8022944,
  author={Wang, Yuanlong and Dong, Daoyi and Qi, Bo and Zhang, Jun and Petersen, Ian R. and Yonezawa, Hidehiro},
  journal={IEEE Trans. Autom. Control}, 
  title={A Quantum Hamiltonian Identification Algorithm: Computational Complexity and Error Analysis}, 
  year={2018},
  volume={63},
  number={5},
  pages={1388-1403},
  doi={10.1109/TAC.2017.2747507}}

@ARTICLE{9026783,
  author={Wang, Yuanlong and Dong, Daoyi and Sone, Akira and Petersen, Ian R. and Yonezawa, Hidehiro and Cappellaro, Paola},
  journal={IEEE Trans. Autom. Control}, 
  title={Quantum Hamiltonian Identifiability via a Similarity Transformation Approach and Beyond}, 
  year={2020},
  volume={65},
  number={11},
  pages={4632-4647},
  doi={10.1109/TAC.2020.2973582}}

@article{HOU2017863,
title = {Experimental quantum Hamiltonian identification from measurement time traces},
journal = {Science Bulletin},
volume = {62},
number = {12},
pages = {863-868},
year = {2017},
issn = {2095-9273},
doi = {https://doi.org/10.1016/j.scib.2017.05.013},
url = {https://www.sciencedirect.com/science/article/pii/S2095927317302554},
author = {Shi-Yao Hou and Hang Li and Gui-Lu Long}
}

@article{PhysRevLett.90.193601,
  title = {Ancilla-Assisted Quantum Process Tomography},
  author = {Altepeter, J. B. and Branning, D. and Jeffrey, E. and Wei, T. C. and Kwiat, P. G. and Thew, R. T. and O'Brien, J. L. and Nielsen, M. A. and White, A. G.},
  journal = {Phys. Rev. Lett.},
  volume = {90},
  issue = {19},
  pages = {193601},
  numpages = {4},
  year = {2003},
  month = {May},
  publisher = {American Physical Society},
  doi = {10.1103/PhysRevLett.90.193601},
  url = {https://link.aps.org/doi/10.1103/PhysRevLett.90.193601}
}

@article{PhysRevA.77.032322,
  title = {Quantum-process tomography: Resource analysis of different strategies},
  author = {Mohseni, M. and Rezakhani, A. T. and Lidar, D. A.},
  journal = {Phys. Rev. A},
  volume = {77},
  issue = {3},
  pages = {032322},
  numpages = {15},
  year = {2008},
  month = {Mar},
  publisher = {American Physical Society},
  doi = {10.1103/PhysRevA.77.032322},
  url = {https://link.aps.org/doi/10.1103/PhysRevA.77.032322}
}

@article{
doi:10.1126/science.1162086,
author = {Mirko Lobino  and Dmitry Korystov  and Connor Kupchak  and Eden Figueroa  and Barry C. Sanders  and A. I. Lvovsky },
title = {Complete Characterization of Quantum-Optical Processes},
journal = {Science},
volume = {322},
number = {5901},
pages = {563-566},
year = {2008},
doi = {10.1126/science.1162086},
URL = {https://www.science.org/doi/abs/10.1126/science.1162086}}

@article{PhysRevLett.106.220503,
  title = {Experimental State Tomography of Itinerant Single Microwave Photons},
  author = {Eichler, C. and Bozyigit, D. and Lang, C. and Steffen, L. and Fink, J. and Wallraff, A.},
  journal = {Phys. Rev. Lett.},
  volume = {106},
  issue = {22},
  pages = {220503},
  numpages = {4},
  year = {2011},
  month = {Jun},
  publisher = {American Physical Society},
  doi = {10.1103/PhysRevLett.106.220503},
  url = {https://link.aps.org/doi/10.1103/PhysRevLett.106.220503}
}

@article{RevModPhys.81.299,
  title = {Continuous-variable optical quantum-state tomography},
  author = {Lvovsky, A. I. and Raymer, M. G.},
  journal = {Rev. Mod. Phys.},
  volume = {81},
  issue = {1},
  pages = {299--332},
  numpages = {0},
  year = {2009},
  month = {Mar},
  publisher = {American Physical Society},
  doi = {10.1103/RevModPhys.81.299},
  url = {https://link.aps.org/doi/10.1103/RevModPhys.81.299}
}

@article{PhysRevResearch.4.033220,
  title = {Continuous-variable quantum state tomography of photoelectrons},
  author = {Laurell, H. and Finkelstein-Shapiro, D. and Dittel, C. and Guo, C. and Demjaha, R. and Ammitzb\"oll, M. and Weissenbilder, R. and Neori\ifmmode \check{c}\else \v{c}\fi{}i\ifmmode \acute{c}\else \'{c}\fi{}, L. and Luo, S. and Gisselbrecht, M. and Arnold, C. L. and Buchleitner, A. and Pullerits, T. and L'Huillier, A. and Busto, D.},
  journal = {Phys. Rev. Res.},
  volume = {4},
  issue = {3},
  pages = {033220},
  numpages = {13},
  year = {2022},
  month = {Sep},
  publisher = {American Physical Society},
  doi = {10.1103/PhysRevResearch.4.033220},
  url = {https://link.aps.org/doi/10.1103/PhysRevResearch.4.033220}
}

@article{PhysRevA.108.042430,
  title = {Continuous-variable quantum tomography of high-amplitude states},
  author = {Fedotova, Ekaterina and Kuznetsov, Nikolai and Tiunov, Egor and Ulanov, A. E. and Lvovsky, A. I.},
  journal = {Phys. Rev. A},
  volume = {108},
  issue = {4},
  pages = {042430},
  numpages = {7},
  year = {2023},
  month = {Oct},
  publisher = {American Physical Society},
  doi = {10.1103/PhysRevA.108.042430},
  url = {https://link.aps.org/doi/10.1103/PhysRevA.108.042430}
}

@article{harriman1993husimi,
  title={Husimi representation for stationary states},
  author={Harriman, John E and Casida, Mark E},
  journal={International journal of quantum chemistry},
  volume={45},
  number={3},
  pages={263--294},
  year={1993},
  publisher={Wiley Online Library}
}

@article{PhysRev.40.749,
  title = {On the Quantum Correction For Thermodynamic Equilibrium},
  author = {Wigner, E.},
  journal = {Phys. Rev.},
  volume = {40},
  issue = {5},
  pages = {749--759},
  numpages = {0},
  year = {1932},
  month = {Jun},
  publisher = {American Physical Society},
  doi = {10.1103/PhysRev.40.749},
  url = {https://link.aps.org/doi/10.1103/PhysRev.40.749}
}

@article{PhysRevA.75.042108,
  title = {Diluted maximum-likelihood algorithm for quantum tomography},
  author = {\ifmmode \check{R}\else \v{R}\fi{}eh\'a\ifmmode \check{c}\else \v{c}\fi{}ek, Jaroslav and Hradil, Zden\ifmmode \check{e}\else \v{e}\fi{}k and Knill, E. and Lvovsky, A. I.},
  journal = {Phys. Rev. A},
  volume = {75},
  issue = {4},
  pages = {042108},
  numpages = {5},
  year = {2007},
  month = {Apr},
  publisher = {American Physical Society},
  doi = {10.1103/PhysRevA.75.042108},
  url = {https://link.aps.org/doi/10.1103/PhysRevA.75.042108}
}

@article{PhysRevA.85.052120,
  title = {Adaptive Bayesian quantum tomography},
  author = {Husz\'ar, F. and Houlsby, N. M. T.},
  journal = {Phys. Rev. A},
  volume = {85},
  issue = {5},
  pages = {052120},
  numpages = {5},
  year = {2012},
  month = {May},
  publisher = {American Physical Society},
  doi = {10.1103/PhysRevA.85.052120},
  url = {https://link.aps.org/doi/10.1103/PhysRevA.85.052120}
}

@article{Granade_2016,
doi = {10.1088/1367-2630/18/3/033024},
url = {https://dx.doi.org/10.1088/1367-2630/18/3/033024},
year = {2016},
month = {mar},
publisher = {IOP Publishing},
volume = {18},
number = {3},
pages = {033024},
author = {Granade, Christopher and Combes, Joshua and Cory, D G},
title = {Practical Bayesian tomography},
journal = {New J. Phys.}
}

@article{Chapman:22,
author = {Joseph C. Chapman and Joseph M. Lukens and Bing Qi and Raphael C. Pooser and Nicholas A. Peters},
journal = {Opt. Exp.},
number = {9},
pages = {15184--15200},
publisher = {Optica Publishing Group},
title = {Bayesian homodyne and heterodyne tomography},
volume = {30},
month = {Apr},
year = {2022},
url = {https://opg.optica.org/oe/abstract.cfm?URI=oe-30-9-15184},
doi = {10.1364/OE.456597},
}

@article{PhysRevLett.127.140502,
  title = {Quantum State Tomography with Conditional Generative Adversarial Networks},
  author = {Ahmed, Shahnawaz and S\'anchez Mu\~noz, Carlos and Nori, Franco and Kockum, Anton Frisk},
  journal = {Phys. Rev. Lett.},
  volume = {127},
  issue = {14},
  pages = {140502},
  numpages = {8},
  year = {2021},
  month = {Sep},
  publisher = {American Physical Society},
  doi = {10.1103/PhysRevLett.127.140502},
  url = {https://link.aps.org/doi/10.1103/PhysRevLett.127.140502}
}

@article{doi:10.1142/S0219749921400050,
author = {Chen, Chuangtao and He, Zhimin and Huang, Zhiming and Situ, Haozhen},
title = {Reconstructing a quantum state with a variational autoencoder},
journal = {International Journal of Quantum Information},
volume = {19},
number = {08},
pages = {2140005},
year = {2021},
doi = {10.1142/S0219749921400050},

URL = { 
    
        https://doi.org/10.1142/S0219749921400050
    
    

}
}

@article{PhysRevApplied.18.044041,
  title = {Simple, Reliable, and Noise-Resilient Continuous-Variable Quantum State Tomography with Convex Optimization},
  author = {Strandberg, Ingrid},
  journal = {Phys. Rev. Appl.},
  volume = {18},
  issue = {4},
  pages = {044041},
  numpages = {12},
  year = {2022},
  month = {Oct},
  publisher = {American Physical Society},
  doi = {10.1103/PhysRevApplied.18.044041},
  url = {https://link.aps.org/doi/10.1103/PhysRevApplied.18.044041}
}

@article{PhysRevResearch.6.033034,
  title = {Efficient factored gradient descent algorithm for quantum state tomography},
  author = {Wang, Yong and Liu, Lijun and Cheng, Shuming and Li, Li and Chen, Jie},
  journal = {Phys. Rev. Res.},
  volume = {6},
  issue = {3},
  pages = {033034},
  numpages = {11},
  year = {2024},
  month = {Jul},
  publisher = {American Physical Society},
  doi = {10.1103/PhysRevResearch.6.033034},
  url = {https://link.aps.org/doi/10.1103/PhysRevResearch.6.033034}
}

@article{WU2024109169,
title = {Supervised training of neural-network quantum states for the next-nearest neighbor Ising model},
journal = {Computer Physics Communications},
volume = {300},
pages = {109169},
year = {2024},
issn = {0010-4655},
doi = {https://doi.org/10.1016/j.cpc.2024.109169},
url = {https://www.sciencedirect.com/science/article/pii/S0010465524000924},
author = {Zheyu Wu and Remmy Zen and Heitor {P. Casagrande} and Dario Poletti and Stéphane Bressan}
}

@book{boyd2004convex,
  title={Convex Optimization},
  author={Boyd, Stephen and Vandenberghe, Lieven},
  year={2004},
  publisher={Cambridge University Press},
  url={https://web.stanford.edu/~boyd/cvxbook/}
}

@article{doi:10.1142/S2251158X12000069,
author = {Curtright, Thomas L. and Zachos, Cosmas K.},
title = {Quantum Mechanics in Phase Space},
journal = {Asia Pacific Physics Newsletter},
volume = {01},
number = {01},
pages = {37-46},
year = {2012},
doi = {10.1142/S2251158X12000069},

URL = { 
    
        https://doi.org/10.1142/S2251158X12000069
    
    

}
}

@book{Hall2013,
  author    = {Brian C. Hall},
  title     = {Quantum Theory for Mathematicians},
  series    = {Graduate Texts in Mathematics},
  volume    = {267},
  year      = {2013},
  publisher = {Springer},
  isbn      = {978-1-4614-7115-8},
  doi       = {10.1007/978-1-4614-7116-5},
  url       = {https://link.springer.com/book/10.1007/978-1-4614-7116-5}
}

@inproceedings{renyi1961measures,
  title={On measures of entropy and information},
  author={R{\'e}nyi, Alfr{\'e}d},
  booktitle={Proceedings of the fourth Berkeley symposium on mathematical statistics and probability, volume 1: contributions to the theory of statistics},
  volume={4},
  pages={547--562},
  year={1961},
  organization={University of California Press}
}

@article{onoe2023direct,
  title={Direct measurement of the Husimi-Q function of the electric-field in the time-domain},
  author={Onoe, Sho and Virally, St{\'e}phane and Seletskiy, Denis V},
  journal={arXiv preprint arXiv:2307.13088},
  year={2023}
}

@Article{Kirchmair2013,
author={Kirchmair, Gerhard
and Vlastakis, Brian
and Leghtas, Zaki
and Nigg, Simon E.
and Paik, Hanhee
and Ginossar, Eran
and Mirrahimi, Mazyar
and Frunzio, Luigi
and Girvin, S. M.
and Schoelkopf, R. J.},
title={Observation of quantum state collapse and revival due to the single-photon Kerr effect},
journal={Nature},
year={2013},
month={Mar},
day={01},
volume={495},
number={7440},
pages={205-209},
issn={1476-4687},
doi={10.1038/nature11902},
url={https://doi.org/10.1038/nature11902}
}

@article{PhysRevA.60.674,
  title = {Direct measurement of the Wigner function by photon counting},
  author = {Banaszek, K. and Radzewicz, C. and W\'odkiewicz, K. and Krasi\ifmmode \acute{n}\else \'{n}\fi{}ski, J. S.},
  journal = {Phys. Rev. A},
  volume = {60},
  issue = {1},
  pages = {674--677},
  numpages = {0},
  year = {1999},
  month = {Jul},
  publisher = {American Physical Society},
  doi = {10.1103/PhysRevA.60.674},
  url = {https://link.aps.org/doi/10.1103/PhysRevA.60.674}
}

@article{10.1214/aop/1176996454,
author = {I. Csiszar},
title = {{$I$-Divergence Geometry of Probability Distributions and Minimization Problems}},
volume = {3},
journal = {The Annals of Probability},
number = {1},
publisher = {Institute of Mathematical Statistics},
pages = {146 -- 158},
year = {1975},
doi = {10.1214/aop/1176996454},
URL = {https://doi.org/10.1214/aop/1176996454}
}

@inproceedings{shalev2009stochastic,
  title={Stochastic Convex Optimization.},
  author={Shalev-Shwartz, Shai and Shamir, Ohad and Srebro, Nathan and Sridharan, Karthik},
  booktitle={COLT},
  volume={2},
  number={4},
  pages={5},
  year={2009}
}

@article{bubeck2015convex,
  title={Convex optimization: Algorithms and complexity},
  author={Bubeck, S{\'e}bastien and others},
  journal={Foundations and Trends{\textregistered} in Machine Learning},
  volume={8},
  number={3-4},
  pages={231--357},
  year={2015},
  publisher={Now Publishers, Inc.}
}

@inproceedings{xu2017stochastic,
  title={Stochastic convex optimization: Faster local growth implies faster global convergence},
  author={Xu, Yi and Lin, Qihang and Yang, Tianbao},
  booktitle={International Conference on Machine Learning},
  pages={3821--3830},
  year={2017},
  organization={PMLR}
}

@article{KREER199465,
title = {Analytic birth—death processes: A Hilbert-space approach},
journal = {Stochastic Processes and their Applications},
volume = {49},
number = {1},
pages = {65-74},
year = {1994},
issn = {0304-4149},
doi = {https://doi.org/10.1016/0304-4149(94)90112-0},
url = {https://www.sciencedirect.com/science/article/pii/0304414994901120},
author = {Markus Kreer}
}

@article{baker1961pade,
  title={The pad{\'e} approximant},
  author={Baker Jr, George A and Gammel, John L},
  journal={Journal of Mathematical Analysis and Applications},
  volume={2},
  number={1},
  pages={21--30},
  year={1961},
  publisher={Elsevier}
}

@article{PhysRevA.41.2645,
  title = {Properties of displaced number states},
  author = {de Oliveira, F. A. M. and Kim, M. S. and Knight, P. L. and Bu\ifmmode \check{z}\else \v{z}\fi{}ek, V.},
  journal = {Phys. Rev. A},
  volume = {41},
  issue = {5},
  pages = {2645--2652},
  numpages = {0},
  year = {1990},
  month = {Mar},
  publisher = {American Physical Society},
  doi = {10.1103/PhysRevA.41.2645},
  url = {https://link.aps.org/doi/10.1103/PhysRevA.41.2645}
}

@incollection{absil2009optimization,
  title={Optimization algorithms on matrix manifolds},
  author={Absil, P-A and Mahony, Robert and Sepulchre, Rodolphe},
  booktitle={Optimization Algorithms on Matrix Manifolds},
  year={2009},
  publisher={Princeton University Press}
}

@book{sato2021riemannian,
  title={Riemannian optimization and its applications},
  author={Sato, Hiroyuki},
  volume={670},
  year={2021},
  publisher={Springer}
}

@article{vapnik1991principles,
  title={Principles of risk minimization for learning theory},
  author={Vapnik, Vladimir},
  journal={Advances in neural information processing systems},
  volume={4},
  year={1991}
}

@article{diamond2016cvxpy,
  author  = {Steven Diamond and Stephen Boyd},
  title   = {{CVXPY}: {A} {P}ython-embedded modeling language for convex optimization},
  journal = {Journal of Machine Learning Research},
  year    = {2016},
  volume  = {17},
  number  = {83},
  pages   = {1--5},
}

@article{Gonçalves03032016,
author = {D.S. Gonçalves and M.A. Gomes-Ruggiero and C. Lavor and},
title = {A projected gradient method for optimization over density matrices},
journal = {Optimization Methods and Software},
volume = {31},
number = {2},
pages = {328--341},
year = {2016},
publisher = {Taylor \& Francis},
doi = {10.1080/10556788.2015.1082105},


URL = { 
    
        https://doi.org/10.1080/10556788.2015.1082105
    
    

},


}

@Article{Bolduc2017,
author={Bolduc, Eliot
and Knee, George C.
and Gauger, Erik M.
and Leach, Jonathan},
title={Projected gradient descent algorithms for quantum state tomography},
journal={npj Quantum Information},
year={2017},
month={Oct},
day={19},
volume={3},
number={1},
pages={44},
issn={2056-6387},
doi={10.1038/s41534-017-0043-1},
url={https://doi.org/10.1038/s41534-017-0043-1}
}

@article{10.1214/10-STS351,
author = {Christian Robert and George Casella},
title = {{A Short History of Markov Chain Monte Carlo: Subjective Recollections from Incomplete Data}},
volume = {26},
journal = {Statistical Science},
number = {1},
publisher = {Institute of Mathematical Statistics},
pages = {102 -- 115},
year = {2011},
doi = {10.1214/10-STS351},
URL = {https://doi.org/10.1214/10-STS351}
}

@article{alizadeh2003second,
  title={Second-order cone programming},
  author={Alizadeh, Farid and Goldfarb, Donald},
  journal={Mathematical programming},
  volume={95},
  number={1},
  pages={3--51},
  year={2003},
  publisher={Citeseer}
}

@article{ocpb:16,
    author       = {Brendan O'Donoghue and Eric Chu and Neal Parikh and Stephen Boyd},
    title        = {Conic Optimization via Operator Splitting and Homogeneous Self-Dual Embedding},
    journal      = {Journal of Optimization Theory and Applications},
    month        = {June},
    year         = {2016},
    volume       = {169},
    number       = {3},
    pages        = {1042-1068},
    url          = {http://stanford.edu/~boyd/papers/scs.html},
}

@ARTICLE{2020NumPy-Array,
  author  = {Harris, Charles R. and Millman, K. Jarrod and
            van der Walt, Stéfan J and Gommers, Ralf and
            Virtanen, Pauli and Cournapeau, David and
            Wieser, Eric and Taylor, Julian and Berg, Sebastian and
            Smith, Nathaniel J. and Kern, Robert and Picus, Matti and
            Hoyer, Stephan and van Kerkwijk, Marten H. and
            Brett, Matthew and Haldane, Allan and
            Fernández del Río, Jaime and Wiebe, Mark and
            Peterson, Pearu and Gérard-Marchant, Pierre and
            Sheppard, Kevin and Reddy, Tyler and Weckesser, Warren and
            Abbasi, Hameer and Gohlke, Christoph and
            Oliphant, Travis E.},
  title   = {Array programming with {NumPy}},
  journal = {Nature},
  year    = {2020},
  volume  = {585},
  pages   = {357–362},
  doi     = {10.1038/s41586-020-2649-2}
}

@software{jax2018github,
  author = {James Bradbury and Roy Frostig and Peter Hawkins and Matthew James Johnson and Chris Leary and Dougal Maclaurin and George Necula and Adam Paszke and Jake Vander{P}las and Skye Wanderman-{M}ilne and Qiao Zhang},
  title = {{JAX}: composable transformations of {P}ython+{N}um{P}y programs},
  url = {http://github.com/jax-ml/jax},
  version = {0.3.13},
  year = {2018},
}

@article{Abril-Pla_PyMC_a_modern_2023,
author = {Abril-Pla, Oriol and Andreani, Virgile and Carroll, Colin and Dong, Larry and Fonnesbeck, Christopher J. and Kochurov, Maxim and Kumar, Ravin and Lao, Junpeng and Luhmann, Christian C. and Martin, Osvaldo A. and Osthege, Michael and Vieira, Ricardo and Wiecki, Thomas and Zinkov, Robert},
doi = {10.7717/peerj-cs.1516},
journal = {PeerJ Comput. Sci.},
month = sep,
title = {{PyMC: a modern, and comprehensive probabilistic programming framework in Python}},
url = {https://peerj.com/articles/cs-1516},
volume = {9},
year = {2023}
}

@article{PRXQuantum.6.010303,
  title = {Demonstrating Efficient and Robust Bosonic State Reconstruction via Optimized Excitation Counting},
  author = {Krisnanda, Tanjung and Fontaine, Clara Yun and Copetudo, Adrian and Song, Pengtao and Lee, Kai Xiang and Huang, Ni-Ni and Valadares, Fernando and Liew, Timothy C.H. and Gao, Yvonne Y.},
  journal = {PRX Quantum},
  volume = {6},
  issue = {1},
  pages = {010303},
  numpages = {20},
  year = {2025},
  month = {Jan},
  publisher = {American Physical Society},
  doi = {10.1103/PRXQuantum.6.010303},
  url = {https://link.aps.org/doi/10.1103/PRXQuantum.6.010303}
}

@misc{sunada2024efficienttomographymicrowavephotonic,
      title={Efficient tomography of microwave photonic cluster states}, 
      author={Yoshiki Sunada and Shingo Kono and Jesper Ilves and Takanori Sugiyama and Yasunari Suzuki and Tsuyoshi Okubo and Shuhei Tamate and Yutaka Tabuchi and Yasunobu Nakamura},
      year={2024},
      eprint={2410.03345},
      archivePrefix={arXiv},
      primaryClass={quant-ph},
      url={https://arxiv.org/abs/2410.03345}, 
}

@Article{O’Sullivan2025,
author={O'Sullivan, James
and Reuer, Kevin
and Grigorev, Aleksandr
and Dai, Xi
and Hern{\'a}ndez-Ant{\'o}n, Alonso
and Mu{\~{n}}oz-Arias, Manuel H.
and Hellings, Christoph
and Flasby, Alexander
and Colao Zanuz, Dante
and Besse, Jean-Claude
and Blais, Alexandre
and Malz, Daniel
and Eichler, Christopher
and Wallraff, Andreas},
title={Deterministic generation of two-dimensional multi-photon cluster states},
journal={Nat. Commun.},
year={2025},
month={Jul},
day={01},
volume={16},
number={1},
pages={5505},
issn={2041-1723},
doi={10.1038/s41467-025-60472-3},
url={https://doi.org/10.1038/s41467-025-60472-3}
}

@article{Lange2023adaptivequantum,
  doi = {10.22331/q-2023-10-09-1129},
  url = {https://doi.org/10.22331/q-2023-10-09-1129},
  title = {Adaptive {Q}uantum {S}tate {T}omography with {A}ctive {L}earning},
  author = {Lange, Hannah and Kebri{\v{c}}, Matja{\v{z}} and Buser, Maximilian and Schollw{\"{o}}ck, Ulrich and Grusdt, Fabian and Bohrdt, Annabelle},
  journal = {{Quantum}},
  issn = {2521-327X},
  publisher = {{Verein zur F{\"{o}}rderung des Open Access Publizierens in den Quantenwissenschaften}},
  volume = {7},
  pages = {1129},
  month = oct,
  year = {2023}
}

@Article{Tripathi2025,
author={Tripathi, Vinay
and Kowsari, Daria
and Saurav, Kumar
and Zhang, Haimeng
and Levenson-Falk, Eli M.
and Lidar, Daniel A.},
title={Benchmarking Quantum Gates and Circuits},
journal={Chemical Reviews},
year={2025},
month={Jun},
day={25},
publisher={American Chemical Society},
volume={125},
number={12},
pages={5745-5775},
issn={0009-2665},
doi={10.1021/acs.chemrev.4c00870},
url={https://doi.org/10.1021/acs.chemrev.4c00870}
}

@Article{Yang2025,
author={Yang, Jiaying
and Strandberg, Ingrid
and Vivas-Via{\~{n}}a, Alejandro
and Gaikwad, Akshay
and Castillo-Moreno, Claudia
and Kockum, Anton Frisk
and Ullah, Muhammad Asad
and Mu{\~{n}}oz, Carlos S{\'a}nchez
and Eriksson, Axel Martin
and Gasparinetti, Simone},
title={Entanglement of photonic modes from a continuously driven two-level system},
journal={npj Quantum Information},
year={2025},
month={Apr},
day={28},
volume={11},
number={1},
pages={69},
issn={2056-6387},
doi={10.1038/s41534-025-00995-1},

}

@Article{Guo2024,
author={Guo, Yuchen
and Yang, Shuo},
title={Quantum state tomography with locally purified density operators and local measurements},
journal={Commun. Phys.},
year={2024},
month={Oct},
day={06},
volume={7},
number={1},
pages={322},
issn={2399-3650},
doi={10.1038/s42005-024-01813-4},

}

@article{PhysRevA.111.022601,
  title = {Reconstruction of quantum states by applying an analytical optimization model},
  author = {Prasad, Rohit and Ghosh, Pratyay and Thomale, Ronny and Huber-Loyola, Tobias},
  journal = {Phys. Rev. A},
  volume = {111},
  issue = {2},
  pages = {022601},
  numpages = {6},
  year = {2025},
  month = {Feb},
  publisher = {American Physical Society},
  doi = {10.1103/PhysRevA.111.022601},

}

@Article{Kawasaki2024,
author={Kawasaki, Akito
and Ide, Ryuhoh
and Brunel, Hector
and Suzuki, Takumi
and Nehra, Rajveer
and Nakashima, Katsuki
and Kashiwazaki, Takahiro
and Inoue, Asuka
and Umeki, Takeshi
and China, Fumihiro
and Yabuno, Masahiro
and Miki, Shigehito
and Terai, Hirotaka
and Yamashima, Taichi
and Sakaguchi, Atsushi
and Takase, Kan
and Endo, Mamoru
and Asavanant, Warit
and Furusawa, Akira},
title={Broadband generation and tomography of non-Gaussian states for ultra-fast optical quantum processors},
journal={Nat. Commun.},
year={2024},
month={Nov},
day={01},
volume={15},
number={1},
pages={9075},
issn={2041-1723},
doi={10.1038/s41467-024-53408-w},
url={https://doi.org/10.1038/s41467-024-53408-w}
}

@article{odonoghue:21,
    author       = {Brendan O'Donoghue},
    title        = {Operator Splitting for a Homogeneous Embedding of the Linear Complementarity Problem},
    journal      = {{SIAM} Journal on Optimization},
    month        = {August},
    year         = {2021},
    volume       = {31},
    issue        = {3},
    pages        = {1999-2023},
}

@article{aa2020,
  title={Globally Convergent {type--I} {A}nderson Acceleration for Non-Smooth Fixed-Point Iterations},
  author={Junzi Zhang and Brendan O'Donoghue and Stephen Boyd},
  journal={{SIAM} Journal on Optimization},
  volume={30},
  number={4},
  pages={3170--3197},
  year={2020}
}

@article{bauschke2017douglas,
  title={On the Douglas--Rachford algorithm},
  author={Bauschke, Heinz H and Moursi, Walaa M},
  journal={Mathematical Programming},
  volume={164},
  number={1},
  pages={263--284},
  year={2017},
  publisher={Springer}
}

@misc{gaikwad2025gradientdescentmethodsfastquantum,
      title={Gradient-descent methods for fast quantum state tomography}, 
      author={Akshay Gaikwad and Manuel Sebastian Torres and Shahnawaz Ahmed and Anton Frisk Kockum},
      year={2025},
      eprint={2503.04526},
      archivePrefix={arXiv},
      primaryClass={quant-ph},
      url={https://arxiv.org/abs/2503.04526}, 
}

@article{moler2003nineteen,
  title={Nineteen dubious ways to compute the exponential of a matrix, twenty-five years later},
  author={Moler, Cleve and Van Loan, Charles},
  journal={SIAM review},
  volume={45},
  number={1},
  pages={3--49},
  year={2003},
  publisher={SIAM}
}

@article{al2010new,
  title={A new scaling and squaring algorithm for the matrix exponential},
  author={Al-Mohy, Awad H and Higham, Nicholas J},
  journal={SIAM Journal on Matrix Analysis and Applications},
  volume={31},
  number={3},
  pages={970--989},
  year={2010},
  publisher={SIAM}
}

@article{mccandlish2018empirical,
  title={An empirical model of large-batch training},
  author={McCandlish, Sam and Kaplan, Jared and Amodei, Dario and Team, OpenAI Dota},
  journal={arXiv preprint arXiv:1812.06162},
  year={2018}
}

@article{kimble2008quantum,
  title={The quantum internet},
  author={Kimble, H Jeff},
  journal={Nature},
  volume={453},
  number={7198},
  pages={1023--1030},
  year={2008},
  publisher={Nature Publishing Group}
}

@article{carmichael2000statistical,
  title={Statistical Methods in Quantum Optics 1: Master Equations and Fokker-Planck Equations},
  author={Carmichael, Howard J and Scully, Marlan O},
  journal={Physics Today},
  volume={53},
  number={3},
  pages={78--80},
  year={2000},
  publisher={American Institute of Physics}
}

@book{carmichael2008statistical,
  title={Statistical methods in quantum optics 2: Non-classical fields},
  author={Carmichael, Howard J},
  year={2008},
  publisher={Springer}
}

@article{marian1993squeezed,
  title={Squeezed states with thermal noise. I. Photon-number statistics},
  author={Marian, Paulina and Marian, Tudor A},
  journal={Physical Review A},
  volume={47},
  number={5},
  pages={4474},
  year={1993},
  publisher={APS}
}

@article{harris2020array,
 title         = {Array programming with {NumPy}},
 author        = {Charles R. Harris and K. Jarrod Millman and St{\'{e}}fan J.
                 van der Walt and Ralf Gommers and Pauli Virtanen and David
                 Cournapeau and Eric Wieser and Julian Taylor and Sebastian
                 Berg and Nathaniel J. Smith and Robert Kern and Matti Picus
                 and Stephan Hoyer and Marten H. van Kerkwijk and Matthew
                 Brett and Allan Haldane and Jaime Fern{\'{a}}ndez del
                 R{\'{i}}o and Mark Wiebe and Pearu Peterson and Pierre
                 G{\'{e}}rard-Marchant and Kevin Sheppard and Tyler Reddy and
                 Warren Weckesser and Hameer Abbasi and Christoph Gohlke and
                 Travis E. Oliphant},
 year          = {2020},
 month         = sep,
 journal       = {Nature},
 volume        = {585},
 number        = {7825},
 pages         = {357--362},
 doi           = {10.1038/s41586-020-2649-2},
 publisher     = {Springer Science and Business Media {LLC}},
 url           = {https://doi.org/10.1038/s41586-020-2649-2}
}

@unpublished{guilmin2025dynamiqs,
  title  = {Dynamiqs: an open-source Python library for GPU-accelerated and differentiable simulation of quantum systems},
  author = {Pierre Guilmin and Adrien Bocquet and {\'{E}}lie Genois and Daniel Weiss and Ronan Gautier},
  year   = {2025},
  url    = {https://github.com/dynamiqs/dynamiqs}
}

@article{JOHANSSON20121760,
title = {QuTiP: An open-source Python framework for the dynamics of open quantum systems},
journal = {Computer Physics Communications},
volume = {183},
number = {8},
pages = {1760-1772},
year = {2012},
issn = {0010-4655},
doi = {https://doi.org/10.1016/j.cpc.2012.02.021},
url = {https://www.sciencedirect.com/science/article/pii/S0010465512000835},
author = {J.R. Johansson and P.D. Nation and Franco Nori},
}

@article{provaznik2022taming,
  title={Taming numerical errors in simulations of continuous variable non-Gaussian state preparation},
  author={Provazn{\'\i}k, Jan and Filip, Radim and Marek, Petr},
  journal={Scientific Reports},
  volume={12},
  number={1},
  pages={16574},
  year={2022},
  publisher={Nature Publishing Group UK London}
}

@article{Littlebaker,
  title        = {cyrus-baker/efficient-bosonic-tomography},
  author       = {cyrus-baker},
journal={GitHub}, 
  year         = 2025,
  note         = {\url{https://github.com/cyrus-baker/efficient-bosonic-tomography.git} }
}

\end{document}